\documentclass[12pt]{article}

\usepackage[x11names]{xcolor}
\usepackage{tikz}
\usepackage{tikz-cd}
\tikzcdset{scale cd/.style={every label/.append style={scale=#1},    cells={nodes={scale=#1}}}}
\usepackage{cite}
\usepackage{comment}
\usepackage{hyperref}
\usetikzlibrary{decorations.pathreplacing,decorations.markings}
\usepgflibrary{arrows}
\usepgflibrary[arrows]
\usetikzlibrary{arrows}
\usetikzlibrary{positioning}
\usetikzlibrary{cd}
\usetikzlibrary{external} 
\usepackage{graphicx}
\usepackage{amsmath}
\usepackage{amssymb}
\usepackage{yhmath}
\usepackage{mathdots}
\usepackage{MnSymbol}
\usepackage{ascmac}
\usepackage{todonotes}
\usepackage{ulem}

\usepackage{subsupscripts}
\usepackage{braket}
\usepackage{amsthm}
\usepackage{xcolor}
\usepackage{subcaption}
\usepackage{bbm}
\usepackage{todonotes}
\usepackage{cancel}

\makeatletter

\@addtoreset{equation}{section}
\makeatother
\newcommand{\ba}{\begin{eqnarray}}
\newcommand{\ea}{\end{eqnarray}}

\newcommand{\cmt}[1]{\textbf{\textcolor{blue}{#1}}}

\usetikzlibrary{decorations.pathreplacing,decorations.markings}
\tikzset{
  on each segment/.style={
    decorate,
    decoration={
      show path construction,
      moveto code={},
      lineto code={
        \path [#1]
        (\tikzinputsegmentfirst) -- (\tikzinputsegmentlast);
      },
      curveto code={
        \path [#1] (\tikzinputsegmentfirst)
        .. controls
        (\tikzinputsegmentsupporta) and (\tikzinputsegmentsupportb)
        ..
        (\tikzinputsegmentlast);
      },
      closepath code={
        \path [#1]
        (\tikzinputsegmentfirst) -- (\tikzinputsegmentlast);
      },
    },
  },
  mid arrow/.style={postaction={decorate,decoration={
        markings,
        mark=at position .5 with {\arrow[#1]{stealth}}
      }}},
}

\tikzstyle{decision} = [diamond, draw, fill=blue!20, 
    text width=4.5em, text badly centered, node distance=3cm, inner sep=0pt]
\tikzstyle{block} = [rectangle, draw, fill=blue!20, 
    text width=8em, text centered, rounded corners, minimum height=2em]
\tikzstyle{line} = [draw, -latex']
\tikzstyle{cloud} = [draw, ellipse,fill=red!20, node distance=3cm,
    minimum height=2em, text centered]

\DeclareMathOperator*{\Span}{\text{Span }}

\DeclareMathOperator{\tr}{\text{tr}}

\newcommand{\superp}[2]{\genfrac{}{}{0pt}{}{#1}{#2}}

 \def\d{\delta}

 \def\p{\partial}
 
 \def\a{\alpha}
 \def\b{\beta}
 \def\g{\gamma}
 \def\d{\delta}
 \def\e{\varepsilon}
 
 \def\th{\theta}

 \def\l{\lambda}

 \def\s{\sigma}
 \def\t{\tau}
 \def\th{\theta}

 \def\D{\Delta}

 \def\O{\Omega}
 \def\o{\omega }
 
\def\CA{{\mathcal{A}}}

\def\CD{{\mathcal{D}}}

\def\CF{{\mathcal{F}}}

\def\CH{{\mathcal{H}}}

\def\CL{{\mathcal{L}}}

\def\CS{{\mathcal{S}}}

\def\CV{{\mathcal{V}}}

\def\CY{{\mathcal{Y}}}

\def\hf{\dfrac{1}{2}}

\def\vphi{\varphi}

\def\CS{\mathcal{S}}

\def\cY{\mathcal{Y}}

\def\rf{\mathfrak{r}}

\def\vac{\emptyset}

\def\mZ{\mathbb{Z}}

\def\mC{\mathbb{C}}

\def\bsx{\boldsymbol{x}}
\def\bsn{\boldsymbol{n}}
\def\bsy{\boldsymbol{y}}
\def\bszeta{\boldsymbol{\zeta}}

\def\gl{\mathfrak{gl}}
\def\sl{\mathfrak{sl}}

\def\mone{\mathbbm{1}}
\def\pdagger{{\prime\dagger}}

\def\bsr{{\boldsymbol{r}}}

\begin{document}
\begin{titlepage}

\vspace*{-2cm}
\vskip 22mm

\begin{center}
{\huge\bf A Calogero model for the non-Abelian\\
\vspace{5mm}
quantum Hall effect}
\end{center}

\vskip 30mm

\begin{center}
{\Large Jean-Emile Bourgine$^\dagger$, Yutaka Matsuo$^\ast$}
\vspace{4mm}

{\em {}$^\dagger$School of Mathematics and Statistics}\\
{\em University of Melbourne}\\
{\em Parkville, Victoria 3010, Australia}\\
\texttt{bourgine@kias.re.kr}

\vspace{4mm}
{\em {}$^\ast$Department of Physics \& Trans-scale Quantum Science Institute}\\{\em \& Mathematics and Informatics Center, The University of Tokyo}\\
{\em Hongo 7-3-1, Bunkyo-ku, Tokyo 113-0033, Japan}\\
\texttt{matsuo@phys.s.u-tokyo.ac.jp}
\vspace{4mm}
\end{center}
\vfill
\begin{abstract}

A model of the non-Abelian fractional quantum Hall effect is obtained from the diagonalization of the matrix model proposed by Dorey, Tong, and Turner in \cite{Dorey2016}. The Hamiltonian is reminiscent of a spin Calogero-Moser model but involves higher-order symmetric representations of the non-Abelian symmetry. We derive the energy spectrum and show that the Hamiltonian has a triangular action on a certain class of wave functions with a free fermion expression. We deduce the expression of the ground states eigenfunctions and show that they solve a Knizhnik-Zamolodchikov equation. Finally, we discuss the emergence of Kac-Moody symmetries in the large $N$ limit using the level-rank duality and confirm the results obtained previously in \cite{Dorey2016}.

\end{abstract}
\vfill
\end{titlepage}

\setcounter{footnote}{0}

\newpage

\section{Introduction}
Quantum systems with topologically protected states called \textit{non-Abelian anyons} are the best candidates for constructing fault-tolerant quantum computing devices \cite{Kitaev1997}. States of this type are encountered in models of the non-Abelian fractional quantum Hall effect (FQHE) \cite{Blok1991}. The FQHE refers to the observation of plateaux in the Hall conductivity of a system of electrons in a strong magnetic field associated with fractional values of the filling factor. The non-Abelian model considers the possibility of assigning spin-like degrees of freedom to electrons to encode various physical properties like the band structure, the presence of higher Landau levels, or a physical spin. This model reveals excited states with non-trivial statistics that exhibit the characteristics of topologically protected states \cite{Blok1991}. While non-Abelian anyons have not been observed yet in condensed matter systems, they have been obtained very recently in quantum processors \cite{Google2023}\cite{iqbal2023creation}. 

Three-dimensional Chern-Simons theory offers competing descriptions of the FQHE at large distances. For instance, Susskind proposed in \cite{Susskind:2001fb} a description of the abelian FQHE using a non-commutative $U(1)$ Chern-Simons theory at level $k$, and recovered in this way the Laughlin wave function corresponding to the filling factor $\nu=1/(k+1)$. Subsequently, Polychronakos introduced a $U(N)$ matrix model as a regularization of Susskind's model to describe the microscopic dynamics of a droplet of $N$ electrons \cite{Polychronakos2001}. In \cite{Tong2003}, building on recent results on the moduli space of vortices \cite{Hanany2003}, Tong re-interpreted Polychronakos's matrix model as a description of vortices in an Abelian commutative Chern-Simons theory. Following this interpretation, Dorey Tong and Turner (DTT) introduced an extension of Polychronakos's matrix model with an additional $U(p)$ symmetry to render the dynamics of vortices in the non-Abelian Chern-Simons theory.

One of the interests of the matrix model is to relate the Chern-Simons theory descriptions to the conformal field theory (CFT) approach to FQHE wave functions initiated by Moore and Read \cite{Moore1991}. Indeed, Polychronakos's $U(N)$ matrix model reduces after diagonalization to a Calogero Hamiltonian, which describes the dynamics of the vortices. The ground state wave function of this model is known to coincide with the Laughlin wave function, which is identified with a conformal block of the rational torus CFT in \cite{Moore1991}. Similarly, DTT recovered from the non-Abelian model the Blok-Wen wave functions identified with the correlators of Wess-Zumino-Witten (WZW) models with $SU(p)_k\times U(1)_{k'}$ Kac-Moody symmetry \cite{Blok1991}\footnote{For a more recent study on the relation between FQHE and Kac-Moody symmetry, see, for example, \cite{schoutens2016simple}\cite{fuji2017non}.}. In fact, the Kac-Moody currents of the CFT can be directly constructed from the matrix model, as shown in \cite{Dorey:2016hoj,Hu:2023eyx}.

The method followed by DTT in \cite{Dorey2016} for the diagonalization and quantization of the matrix model is based on a matrix version of holomorphic coordinates and differs from Polychronakos' approach. In this paper, we revisit the DTT matrix model following Polychronakos's original approach and show that the dynamics of the vortices are governed by a spin Calogero Hamiltonian involving higher-order symmetric representations. To be specific, the order of the $U(p)$ spin representation is identified with the Chern-Simons level $k$. A generating family of wave functions called \textit{wedge states} \cite{DKJM1983}\cite{Uglov:1996np} is introduced, and we prove that the action of the Hamiltonian on these states is triangular. We deduce from this result the spectrum of our model and the expression of the ground state wave functions, recovering the Blok-Wen wave functions in the absence of degeneracies. More generally, all ground state eigenfunctions are shown to obey the Knizhnik-Zamolodchikov (KZ) equation characteristic of WZW conformal blocks. 

At level $k>1$, the wedge states are no longer linearly independent due to Pl\"ucker relations. We solve this problem using a fermionic representation of the wave functions, which highlights the underlying algebraic structure. In the large $N$ limit, the wedge states become the exact eigenstates of the Hamiltonian. The free fermions are identified with the fields of a 2D CFT with the conformal embedding $\widehat{\gl}(pk)_1 \supset \widehat{\sl}(p)_k\otimes \widehat{\sl}(k)_p\otimes \widehat{\mathfrak{u}}(1)$. This type of CFT has been studied previously in the context of level-rank duality \cite{Nakanishi:1990hj}. The Pl\"ucker relations project out the dual Kac-Moody $\widehat{\sl}(k)_p$ algebra, and one recovers the $\widehat{\sl}(p)_k\otimes\widehat{\mathfrak{u}}(1)$ symmetry of the system in agreement with \cite{Dorey2016}. The primary fields of the Kac-Moody algebra, which appears in the KZ equation, are identified with vertex operators. They are used to translate the wedge states into the states of the fermionic Fock space. In this way, the spin Calogero model is directly related to the 2D conformal field theory.


One of our main motivations is the well-known integrability properties of Calogero-type models, which offers the prospect of applying the algebraic methods developed in this context, like the Algebraic Bethe Ansatz \cite{Ferrando2023} or the Lax formalism \cite{Hikami1993}, to the problem of the FQHE. While we do not prove these properties in this paper, we strongly suspect that our Hamiltonian possesses a Yangian symmetry realized in a similar way as in the lower order version studied in \cite{Bernard1994} (see also \cite{Hikami1993} for an approach using the Lax pair). This expectation is consistent with the well-known fact that Yangian invariance generates Kac-Moody symmetries in the large $N$ limit \cite{Bernard:1994wg, Bouwknegt1994,  Bouwknegt1996}. We hope to be able to come back to these important questions in a future publication.

This paper is organized as follows. Section two contains the derivation of the Hamiltonian by diagonalization of the DTT matrix model. The analysis of its spectrum and eigenfunctions is presented in section three, and the free fermion formalism and the large $N$ limit are explained in section four. Technical details have been gathered in the appendix. Appendix A contains the proof of our main results, namely the triangular action of the Hamiltonian and the KZ equation for the ground states eigenfunction. Appendix B presents the eigenfunctions of the Hamiltonian for the simplest case of $N=2$ particles, and Appendix C is a brief reminder of the construction of the Gelfand-Zetlin basis, which applies to the $k=1$ case. Finally, Appendix D presents an explicit check of the conformal embedding by comparing the explicit expression of characters in the case $p=k=2$.


\section{Derivation of the Hamiltonian}
In this section, we first present our quantum model and then its derivation from the diagonalization of the DTT matrix model.

\subsection{Definition of the model}
The system consists of $N$ particles with coordinates $x_a$ carrying a $U(p)$ spin structure defined using the $Np$ oscillators $[\vphi_{i,a},\vphi_{j,b}^\dagger]=\d_{i,j}\d_{a,b}$. In this paper, we take the convention that indices $a,b,c,d\in\{1,\cdots N\}$ label the particles, and $i,j\in\{1,\cdots,p\}$ their spin states. Each particle carries a symmetric representation $[k]$ of $U(p)$ obtained from the action of $k$ creation operators $\vphi_{i_1,a}^\dagger\cdots \vphi_{i_k,a}^\dagger\ket{\vac}$ on the vacuum $\ket{\vac}$ satisfying 
\begin{equation}
\vphi_{i,a}\ket{\vac}=0,\quad \forall i\in\{1,\cdots p\},\quad \forall a\in\{1,\cdots,N\}.
\end{equation} 
In addition to the integers $N,p,k$, the model depends on the parameter $B$ that confines the particles in a harmonic potential. The Hamiltonian of the system reads\footnote{We note some similarity with the Calogero model obtained in \cite{Copland2012} by dimensional reduction of 2d Super Yang-Mills. However, here the interaction term is bosonic, while it is fermionic in \cite{Copland2012}.}
\begin{equation}\label{Hamiltonian}
\CH=\sum_{a=1}^N\left(-\dfrac{\p^2}{\p x_a^2}+B^2 x_a^2\right)+2\sum_{\superp{a,b=1}{a<b}}^N\dfrac{J_{a,b}J_{b,a}}{(x_a-x_b)^2},\quad J_{a,b}=\sum_{i=1}^p\vphi_{i,a}^\dagger\vphi_{i,b}.
\end{equation} 
This model is expected to describe the non-Abelian quantum Hall states with filling factor $\nu=p/(k+p)$. We note that the generators $J_{a,b}$ satisfy the relations of the $\gl(N)$ Lie algebra $[J_{a,b},J_{c,d}]=\d_{a,d}J_{c,b}-\d_{b,c}J_{a,d}$, and the Hamiltonian commutes with the diagonal operators $J_{a,a}$ for $a=1,\cdots,N$. These operators are diagonal on spin states $\vphi_{i_1,a}^\dagger\cdots \vphi_{i_k,a}^\dagger\ket{\vac}$ in the representation $[k]$, with the eigenvalue $k$. The $U(N)$ constraints of our model impose to fix this integer $k$ to be the same for each particle $a$. The Hamiltonian \ref{Hamiltonian} also exhibits a global $U(p)$ invariance which is generated by the operators
\begin{equation}\label{def_Kij}
K_{i,j}=\sum_{a=1}^N\vphi_{i,a}^\dagger\vphi_{j,a},\quad [K_{i,j},K_{k,l}]=\d_{j,k}K_{i,l}-\d_{i,l}K_{k,j}.
\end{equation}

%
%
%

\paragraph{Abelian model} Setting $p=1$, we recover the Abelian model introduced by Polychronakos in \cite{Polychronakos2001} to describe the Laughlin states at filling factor $\nu=1/(k+1)$. In this case, there is no spin structure as every particle carries the fixed spin state $(\vphi_a^\dagger)^k\ket{\vac}$. The action of the interaction term in the Hamiltonian \ref{Hamiltonian} simplifies since
\begin{equation}
J_{a,b}J_{b,a}\prod_{c=1}^N(\vphi_c^\dagger)^k\ket{\vac}=k(k+1)\prod_{c=1}^N(\vphi_c^\dagger)^k\ket{\vac},
\end{equation} 
and we recover the Calogero-Moser Hamiltonian with a Gaussian potential,
\begin{equation}\label{Hamiltonian_p1}
\CH^{(p=1)}=\sum_{a=1}^N\left(-\dfrac{\p^2}{\p x_a^2}+B^2 x_a^2\right)+2k(k+1)\sum_{\superp{a,b=1}{a<b}}^N\dfrac{1}{(x_a-x_b)^2}.
\end{equation} 
This Hamiltonian is known to be integrable \cite{Olshanetsky1983}.

\paragraph{Spin Calogero-Moser} Considering instead the case $k=1$ for arbitrary $p$, the particles carry the fundamental representation of $U(p)$. This is usually implemented by assigning the spin state $\ket{i_a}=\vphi_{i_a,a}^\dagger\ket{\vac}$ to the particle $a$, with the label $i_a\in\{1,\cdots,p\}$. In this case, the interaction term also simplifies as the product $J_{a,b}J_{b,a}$ acting on such spin states can be rewritten using the permutation $P_{a,b}$ exchanging the spin labels $i_a$ and $i_b$,
\begin{equation}
J_{a,b}J_{b,a} \prod_{c=1}^N\vphi_{i_c,a}^\dagger\ket{\vac}=(1+P_{a,b}) \prod_{c=1}^N\vphi_{i_c,a}^\dagger\ket{\vac}.
\end{equation} 
Then, the Hamiltonian \ref{Hamiltonian} reduces to the spin Calogero-Moser Hamiltonian at coupling $\l=1$ with an extra Gaussian potential,
\begin{equation}
\CH^{(k=1)}=\sum_a\left(-\dfrac{\p^2}{\p x_a^2}+B^2 x_a^2\right)+2\sum_{\superp{a,b=1}{a<b}}^N\dfrac{1+P_{a,b}}{(x_a-x_b)^2}.
\end{equation} 
In the absence of the Gaussian term, this model is also a well-known integrable system which exhibits the Yangian symmetry $Y(\sl_p)$ \cite{Bernard1993,Hikami1993,Hikami1993a}.

\paragraph{Remark} 
Another remarkable value, albeit somewhat trivial, is obtained at $k=0$. In this case, particles carry no spin degree of freedom, and the interaction term of the Hamiltonian \ref{Hamiltonian} vanishes. We are left with $N$ decoupled quantum harmonic oscillators. If we look for an antisymmetric wave function so that the true matrix model wave function is symmetric, the ground state of our Hamiltonian is\footnote{Without the anti-symmetrization requirement, we would simply have
\begin{equation}
\Psi_0(\bsx)=e^{-\frac{B}2\sum_ax_a^2},\quad E_0=NB.
\end{equation}}
\begin{equation}
\Psi_0(\bsx)=\D(\bsx)e^{-\frac{B}2\sum_ax_a^2},\quad E_0=N^2B.
\end{equation}
This is the Slater determinant of the wave functions of $N$ independent quantum harmonic oscillators. It coincides with the Laughlin wave function of filling factor $\nu=1$.

\subsection{Diagonalization of the DTT matrix model}
The Hamiltonian \ref{Hamiltonian} has been obtained from the diagonalization of the matrix quantum mechanics proposed in \cite{Dorey2016} to describe the dynamics of vortices in a $U(p)$ Chern-Simons theory. It is a non-Abelian generalization of the quantum Hall droplet matrix model proposed by Polychronakos in \cite{Polychronakos2001}, which was shown in \cite{Tong2003, Tong2015} to also describe the vortices of the $U(1)$ Chern-Simons theory. The non-Abelian matrix model is formulated in terms of dynamical fields consisting in a $N\times N$ complex matrix $Z(t)$ and $p$ $N$-dimensional vectors $\vphi_i(t)$ ($i=1\cdots p$). The matrix $Z(t)$ transforms in the adjoint representation $Z(t)\to U(t)Z(t)U(t)^\dagger$ under $U(N)$ symmetry, and the vectors $\vphi_i(t)$ in the fundamental representation $\vphi_i(t)\to U(t)\vphi_i(t)$. The model also contains a non-dynamical gauge field $\a$ which imposes the (classical) constraint 
\begin{equation}
\frac{B}2[Z,Z^\dagger]+\sum_{i=1}^p\vphi_i\vphi_i^\dagger=(k+p)\mone_N.
\end{equation} 
The action of this model reads \cite{Dorey2016}
\begin{equation}
\CS=\int dt\left[\hf iB\tr(Z^\dagger\CD_tZ)+i\sum_{i=1}^p\vphi_i^\dagger\CD_t\vphi_i-(k+p)\tr\a-\o\tr(Z^\dagger Z)\right],
\end{equation} 
where $\CD_t$ is the covariant derivative
\begin{equation}
\CD_t Z=\p_t Z-i[\a,Z],\quad \CD_t\vphi_i=\p_t\vphi_i-i\a\vphi_i.
\end{equation} 
The model depends on the parameters $B$ (background magnetic field), $k$ (Chern-Simons level), and $\o$ (strength of the harmonic trap).

The complex matrix $Z$ can be decomposed into a sum of Hermitian matrices
\begin{equation}
Z=X_1+iX_2,\quad Z^\dagger=X_1-iX_2,
\end{equation}
and isolating the dependence in the gauge field, the previous action written in a more explicit form,
\begin{align}
\begin{split}\label{action_X}
&\CS=\int dt\left(-\dfrac{B}{2}\tr\left(X_1\dot{X}_2-X_2\dot{X}_1\right)+i\sum_{i=1}^p\vphi_i^\dagger\dot{\vphi}_i-\o\tr(X_1^2+X_2^2)+\tr\left[\a G(X_1,X_2,\vphi_i,\vphi_i^\dagger)\right]\right),\\
&\text{with}\quad G(X_1,X_2,\vphi_i,\vphi_i^\dagger)=-iB[X_1,X_2]+\sum_{i=1}^p\vphi_i\vphi_i^\dagger-(k+p)\mone_N,
\end{split}
\end{align}
and we used the dot notation to indicate the time derivative. When $p=1$, this is precisely the form of the action studied by Polychronakos in \cite{Polychronakos2001}. In contrast with the work of DTT, we will work in the framework where $(X_1, X_2)$ are the canonical conjugate fields, instead of $(Z, Z^\dagger)$. One of our motivations is to investigate the connection with quantum integrable systems. Indeed, the dynamics of the eigenvalues of the matrix $X_1$ is given by the Calogero-Moser Hamiltonian \ref{Hamiltonian_p1} when $p=1$ \cite{Polychronakos2001}. As we will show shortly, in the non-Abelian model, this dynamics is defined by the Hamiltonian \ref{Hamiltonian}, which we suspect is also integrable. A second, more physical, motivation is the interpretation of these eigenvalues as the effective electron coordinates in the quantum droplet model, and so the position of Chern-Simons vortices. In this way, we expect the wave functions derived in the next section to give a more physical description of the corresponding model.

\paragraph{Diagonalization} The hermitian matrix $X_1$ is diagonalized by decomposing $X_1(t)=\O(t)^\dagger x(t)\O(t)$ with $(x)_{a,b}=x_a\d_{a,b}$ a diagonal matrix and $\O\in\text{SU}(N)$ a unitary matrix containing the $N^2-N$ angular degrees of freedom. This decomposition is introduced using the global $U(N)$ symmetry, and we denote with a prime the transformed quantities
\begin{equation}
X_2'=\O X_2\O^\dagger,\quad \vphi_i'=\O\vphi_i,\quad \vphi_i^\pdagger=\vphi_i^\dagger\O^\dagger,\quad \a'=\O\a\O^\dagger,\quad G'=\O G\O^\dagger.
\end{equation} 
After transformation, the dynamical terms in the Lagrangian read
\begin{align}
\begin{split}
&\tr X_1\dot{X_2}=\tr x\dot{X}_2'+\tr L[x,X'_2],\quad\tr \dot{X_1}X_2=\tr \dot{x}X'_2-\tr L[x,X'_2],\quad\vphi_i^\dagger\dot{\vphi}_i=\vphi_i^\pdagger\dot{\vphi}'_i-\vphi_i^\pdagger L\vphi'_i,
\end{split}
\end{align}
where we introduced the anti-hermitian matrix $L=\dot{\O}\O^\dagger=-\O\dot{\O}^\dagger$. Replacing these terms in the action, we find after integration by part
\begin{align}\label{def_G}
\begin{split}
&\CS=\int dt\left(B\tr\dot{x}X'_2+i\sum_{i=1}^p\vphi_i^\pdagger\dot{\vphi'}_i-\o\tr(x^2+(X_2')^2)-B\tr L[x,X'_2]-i\sum_{i=1}^p\vphi_i^\pdagger L{\vphi'}_i+\tr\a'G'\right),\\
&\text{with}\quad G'=-iB[x,X_2']+\sum_{i=1}^p\vphi_i'\vphi_i^\pdagger-(k+p)\mone_N.
\end{split}
\end{align}
From now on, we drop the prime on the variables to lighten the notation. 

In the action \ref{def_G}, the dynamical fields are the eigenvalues $x_a(t)$, the components $\vphi_{i,a}(t)$ of the vector fields, and the angular degrees of freedom contained in $\O(t)$. Following the standard method reviewed in \cite{Klebanov1991}, the anti-hermitian matrix $L$ can be decomposed over the generators of $su(N)$,
\begin{equation}
L=\sum_{a=1}^{N-1} \dot{\o}_a H_a+\sum_{\superp{a,b=1}{a<b}}^N(\dot{\o}_{a,b} T_{a,b}+\dot{\th}_{a,b}\tilde{T}_{a,b}).
\end{equation} 
We denoted the Cartan generators $H_a=i(e_{a,a}-e_{a+1,a+1})$, and the anti-hermitian matrices $T_{a,b}=i(e_{a,b}+e_{b,a})/\sqrt{2}$ and $\tilde{T}_{a,b}=(e_{a,b}-e_{b,a})/\sqrt{2}$ (here $e_{a,b}$ denotes the elements of the canonical basis of GL$(N)$). Using this decomposition, we can compute the momentum associated to each degree of freedom,
\begin{align}
\begin{split}\label{def_p}
&p_a=\dfrac{\d\CL}{\d \dot{x}_a}=BX_{2\ a,a},\quad \pi_{i,a}=\dfrac{\d\CL}{\d \dot{\vphi}_{i,a}}=i\vphi_{i,a}^\dagger,\quad \Pi_a=\dfrac{\d\CL}{\d \dot{\o}_a}=J_{a,a}-J_{a+1,a+1},\\
&\Pi_{a,b}=\dfrac{\d\CL}{\d \dot{\o}_{a,b}}=\dfrac1{\sqrt{2}}\left(J_{a,b}+J_{b,a}-iB(x_a-x_b)(X_{2\ a,b}-X_{2\ b,a})\right),\\
&\tilde{\Pi}_{a,b}=\dfrac{\d\CL}{\d \dot{\th}_{a,b}}=-\dfrac{i}{\sqrt{2}}\left(J_{a,b}-J_{b,a}+iB(x_a-x_b)(X_{2\ a,b}+X_{2\ b,a})\right),
\end{split}
\end{align}
where $J_{a,b}$ is defined in \ref{Hamiltonian}. We deduce the Hamiltonian, which reduces to a potential term since the Lagrangian depends linearly on time derivatives. Setting $\a=0$, we find
\begin{equation}\label{H1}
\CH=\o\tr(x^2+(X_2)^2)=\o\sum_{a=1}^N\left(x_a^2+\dfrac{p_a^2}{B^2}\right)+2\dfrac{\o}{B^2}\sum_{a<b}\dfrac{(J_{a,b}-\Pi_{a,b}^+)(J_{b,a}-\Pi_{a,b}^-)}{(x_a-x_b)^2},
\end{equation} 
with $\Pi^\pm_{a,b}=(\Pi_{a,b}\pm i\tilde{\Pi}_{a,b})/\sqrt{2}$.

\paragraph{Quantization} We now impose the canonical quantization conditions for the angular variables and the vector fields,
\begin{equation}\label{canonical}
[\o_a,\Pi_b]=i\d_{a,b},\quad [\o_{a,b},\Pi_{c,d}]=i\d_{a,c}\d_{b,d},\quad [\vphi_{i,a},\pi_{j,b}]=i\d_{i,j}\d_{a,b}.
\end{equation} 
In particular, $\vphi_{i,a}$ and $\vphi_{i,a}^\dagger$ are conjugate variables, $[\vphi_{i,a},\vphi_{j,b}^\dagger]=\d_{i,j}\d_{a,b}$. However, we note that due to the Vandermonde determinant in the measure coming from the diagonalization $dX_1=d\O dx\D(\bsx)$, the momentum $p_a$ acts on wave functions as (see, e.g., \cite{Klebanov1991})
\begin{equation}
p_a=-i\D(\bsx)^{-1}\p_a\D(\bsx),\quad \D(\bsx)=\prod_{\superp{a,b=1}{a<b}}^N(x_a-x_b).
\end{equation}

After quantization, the constraints $G_{a,b}=0$ must be imposed on physical wave functions. In the canonical variables, these constraints take the form
\begin{equation}
G_{a,b}
=\left\{
\begin{array}{ll}
J_{a,a}-k& (a=b),\\
\Pi_{a,b}^- & (a<b),\\
\Pi_{b,a}^+ & (a>b).
\end{array}
\right.
\end{equation} 
Thus, physical states must obey
\begin{equation}\label{eq:phys}
\Pi_{a}\ket{\text{phys}}=\Pi_{a,b}\ket{\text{phys}}=\tilde{\Pi}_{a,b}\ket{\text{phys}}=0,\quad J_{a,a}\ket{\text{phys}}=k\ket{\text{phys}}.
\end{equation} 
These relations imply that they have no dependence in the angular variables $\o_a$, $\o_{a,b}$, and $\th_{a,b}$. Since $J_{a,a}$ counts a number of modes, the last condition imposes the quantization $k\in\mZ$ of the Chern-Simons level. From the relation $J_{a,b}J_{b,a}+J_{a,a}=J_{b,a}J_{a,b}+J_{b,b}$, we also deduce that $J_{a,b}J_{b,a}\equiv J_{b,a}J_{a,b}$ on physical states. Thus, the action of the Hamiltonian simplifies on the physical states\footnote{We note that the canonical quantization condition implies that the conjugate variables $\vphi_{i,a}$ and $\vphi_{i,a}^\dagger$ both commute with the momenta $\Pi_{a,b}$ and $\tilde{\Pi}_{a,b}$, and so $J_{a,b}$ commutes with $\Pi_{a,b}^\pm$ in \ref{H1}.}
\begin{equation}
\CH=\dfrac{\o}{B^2}\left(\sum_{a=1}^N\left(-\D(\bsx)^{-1}\p_a^2\D(\bsx)+B^2 x_a^2\right)+2\sum_{a<b}\dfrac{J_{a,b}J_{b,a}}{(x_a-x_b)^2}\right),
\end{equation} 
and it indeed reduces to the Hamiltonian given in equ. \ref{Hamiltonian} upon rescaling the energies by a factor $\o^{-1}B^2$, and a conjugation with the Vandermonde $\CH\to \D(\bsx)\CH\D(\bsx)^{-1}$. The latter simply amounts to multiplying the wave functions by the antisymmetric factor $\D(\bsx)$.

\section{Spectrum and eigenfunctions}
\subsection{Abelian model}
Before studying the general case, we recall the results obtained for the Abelian model ($p=1$). As previously observed, the Hamiltonian \ref{Hamiltonian_p1} describes the Calogero-Moser system with coupling $k\in\mZ$ and a Gaussian potential. For this system, the ground state is well-known,\footnote{We discard the wave function singular at coincident points which leads to negative energies for $k>0$,
\begin{equation}
\Psi_0(\bsx)=\prod_{a<b}(x_a-x_b)^{-k}e^{-\frac{B}2\sum_ax_a^2},\quad E_0=NB-kN(N-1)B.
\end{equation}}
\begin{equation}\label{E0}
\Psi_0(\bsx)=\prod_{a<b}(x_a-x_b)^{k+1}e^{-\frac{B}2\sum_ax_a^2},\quad E_0=N^2B+kN(N-1)B.
\end{equation} 
Upon analytic continuation, $\Psi_0(x)$ coincides with the Laughlin wave function with filling factor $\nu=1/(k+1)$.

This model has been studied by Polychronakos in the context of the abelian FQHE in \cite{Polychronakos2001}, and we recall here some of his results. States of the model are labeled by $N$ positive integers $n_a$ satisfying the constraint $n_{a+1}-n_a\geq k+1$, and the corresponding energy is\footnote{To recover our convention, we simply need to replace  $\o=2B$ in the formulas (45) and (47) of \cite{Polychronakos2001}. Comparing instead with the ground state energy found by DTT in \cite{Dorey2016}, we need to discard an overall shift of $N^2B$. We will find the same overall shift of the energies in the non-Abelian model as well.}
\begin{equation}
E_{\bsn}=B\sum_{a=1}^N(2n_a+1).
\end{equation}
For the ground state, the labels take the values $n_a=(a-1)(k+1)$ which gives precisely the energy $E_0$ written in \ref{E0}. To count the excited states, we can introduce the labels $r_a=n_{a+1}-n_a-(k+1)$ with $a=1,\cdots N-1$, together with $r_0=n_1$, for which the previous condition becomes $r_a\geq 0$. In these variables, the energy reads
\begin{equation}
E_{\bsn}=E_0+2B\sum_{a=1}^N ar_{N-a}.
\end{equation} 
We can interpret the variables $r_{N-a}$ as counting the multiplicities of the columns of height $a$ of a Young diagram with a total of $\sum_a ar_{N-a}$ boxes, each column being restricted to contain at most $N$ boxes. In the large $N$ limit, this restriction drops, and the corresponding character is simply the $u(1)$ WZW character
\begin{equation}\label{U(1)character}
\chi(q)=\sum_{n=0}^\infty p(n)q^{E_0+2Bn}=\dfrac{q^{E_0}}{\prod_{j=1}^\infty(1-q^{2Bj})},
\end{equation} 
where $p(n)$ is the number of partitions of $n$.

\subsection{Non-Abelian model}
In the following, for the sake of simplicity, we introduce a polynomial representation of the Fock space associated to the variables $\varphi_{i,a}$,
\begin{equation}
	\varphi^\dagger_{i_1,a_1}\cdots \varphi^{\dagger}_{i_n,a_n}|0\rangle \quad \rightarrow \quad y_{i_1,a_1}\cdots y_{i_n,a_n}.
\end{equation}
In this representation, the operators $\varphi_{i,a}$ act as the derivatives $\partial_{i,a}=\p/\p y_{i,a}$.
The $\mathfrak{gl}(N)$ generators are represented by
\begin{equation}
	J_{a,b}=\sum_{i=1}^p y_{i,a}\partial_{i,b}\,.
\end{equation}
The condition (\ref{eq:phys}) implies that physical states are described by a homogeneous polynomial of degree $k$ in $y_{i, a}$ for each particle $a=1,\cdots, N$.

In this section, we derive the eigenvalues and study the eigenfunctions of the Hamiltonian (\ref{Hamiltonian}). To simplify the expression of wave functions, it is convenient to modify the Hamiltonian and absorb the Gaussian part,
\begin{equation}\label{Htilde_m}
	\tilde{\CH}=e^{\frac{B}2 \sum_a x_a^2} \CH e^{-\frac{B}2 \sum_a x_a^2}=\sum_a\left(
	-\frac{\partial^2}{\partial x_a^2} +2Bx_a\frac{\partial}{\partial x_a}
	\right)+NB+\sum_{a\neq b} \frac{J_{a,b}J_{b,a}}{(x_a-x_b)^2}.
\end{equation}
We further use the rescaling of coordinates $x_a\to B^{-1/2}x_a$ and energies $\tilde{\CH}\to B^{-1}\tilde{\CH}$ to set $B=1$. 

The diagonalization of the Hamiltonian \ref{Htilde_m} is treated explicitly in appendix \ref{a:N=2} in the case of two particles (i.e. $N=2$) with $k,p$ generic. The general treatment for an arbitrary number of particles given below follows from a careful analysis of specific cases at lower $N$ which are not presented in this paper to avoid redundancy.

\paragraph{Wedge states}
In order to write down the wave functions of the Hamiltonian \ref{Htilde_m}, we introduce the wedge product notation for the following $N\times N$ determinants,
\begin{equation}
    [y_{i_1}x^{n_1}\wedge \cdots \wedge y_{i_N}x^{n_N}]=\det\left(\begin{array}{ccc}
    y_{i_1,1}(x_1)^{n_1} & \cdots & y_{i_N,1} (x_1)^{n_N}\\
    \vdots & & \vdots\\
    y_{i_1,N}(x_N)^{n_1} & \cdots & y_{i_N,N}(x_N)^{n_N}
    \end{array}
    \right).
\end{equation}
It satisfies the linearity and the anti-symmetry properties,
\begin{align}
    & [\cdots\wedge (\lambda_1 u_1 +\lambda_2 u_2)\wedge \cdots] =\lambda_1[\cdots\wedge u_1 \wedge \cdots]+\lambda_2[\cdots \wedge u_2 \wedge \cdots],\quad
    \lambda_1,\lambda_2\in \mathbb{C}\\
    &[\cdots\wedge \underset{\hat{a}}{u_1} \wedge \cdots \wedge\underset{\hat{b}}{u_2} \wedge \cdots ]=-[\cdots\wedge \underset{\hat{a}}{u_2} \wedge \cdots \wedge \underset{\hat{b}}{u_1} \wedge \cdots ]\,.\label{anti-symmetry}
\end{align}
The second property implies that $[\cdots\wedge \underset{\hat{a}}{u} \wedge \cdots \wedge\underset{\hat{b}}{u} \wedge \cdots ]=0$.

The elementary blocks for constructing wave functions are the elements of the following family of determinants indexed by $N$-tuple integers $\boldsymbol{r}=(r_1,\cdots, r_N)$,
\begin{equation}\label{Yl}
\cY_{\boldsymbol{r}}(\bsx,\bsy)=\det_{a,b}(y_{i_b,a}x_a^{n_b})=[y_{i_1}x^{n_1}\wedge \cdots \wedge y_{i_N}x^{n_N}].
\end{equation}
We call these determinants \textit{wedge states}. The components $r_a$ of the index $\bsr$ are related to the parameters $(n_a,i_a)$ of the particle $a$ by Euclidean division $r_a=pn_a+i_a-1$, $1\leq i_a\leq p$. This notation comes from the fact that monomials $y_{i,a} (x_a)^n$ can be represented as a power of another variable $z_a^{r}$, where $n$ and $i-1$ are respectively the quotient and remainder of the Euclidean division of $r=pn+i-1$ by $p$. Such a notation is used, e.g., in \cite{Uglov1997}.
In these variables, the determinants $\cY_{\boldsymbol{r}}(\bsx,\bsy)$ take the simpler forms
\begin{equation}
    \cY_{\boldsymbol{r}}(\bsx,\bsy)=\det_{a,b} z_a^{r_b}=[z^{r_1}\wedge \cdots \wedge z^{r_N}].
\end{equation}
By definition, the determinants vanish if $r_a=r_b$ for $a\neq b$, and up to permutation we can assume $0\leq r_1<r_2<\cdots<r_N$. Finally, we mention the Pl\"ucker relations for bilinear combinations of the wedge products states\footnote{See, e.g., \cite{Griffiths-Harris}, p211. In the context of the soliton equation, it appeared as the Hirota bilinear equation for the tau function \cite{DKJM1983}.},
\begin{equation}\label{Plucker}
    \sum_{a=0}^N (-1)^a [z^{r_1}\wedge\cdots \wedge z^{r_{N-1}}\wedge z^{s_a}][z^{s_0}\wedge\cdots \wedge \cancel{z^{s_a}}\wedge \cdots \wedge z^{s_N}]=0,
\end{equation}
for arbitrary pair $\boldsymbol{r}=(r_1,\cdots, r_{N-1})$ and $\boldsymbol{s}=(s_0,\cdots,s_N)$. It produces nontrivial quadratic relations for arbitrary choice of $\boldsymbol{r}$ and $\boldsymbol{s}$ if $\boldsymbol{r}\not\subset \boldsymbol{s}$.

\paragraph{Eigenfunctions as a product of wedge states} To label the eigenstates, we introduce a set of $k$ sequences of $N$ non-negative integers,
\begin{align}
    \mathfrak{r}&=\left\{\boldsymbol{r}^{(1)},\cdots, \boldsymbol{r}^{(k)}    \right\}\\
    \boldsymbol{r}^{(\alpha)}&=\{r^{(\alpha)}_1,\cdots , r^{(\alpha)}_N\},\quad 0\leq r^{(\alpha)}_1<r^{(\alpha)}_2<\cdots<r^{(\alpha)}_N\,,\quad
    \alpha=1,\cdots, k\,.
\end{align}
We suppose that the re-ordering of $\boldsymbol{r}^{(\alpha)}$ in $\mathfrak{r}$ does not give a new element. Let $\mathfrak{R}$ be the set of all possible $\mathfrak{r}$. For each element $\mathfrak{r}\in \mathfrak{R}$, we define a function of the variables $x_a,y_{i,a}$ 
\begin{align}\label{psil}
	\psi_{\mathfrak{r}}(\bsx,\bsy)&= \D(\bsx) \prod_{\alpha=1}^k \cY_{\boldsymbol{r}^{(\alpha)}}(\bsx,\bsy)\,.
\end{align}
It can also be seen as an element in $\mC[z_1,\cdots, z_N]$ using the previous translation rule. We define the level of $\psi_{\mathfrak{r}}(\bsx,\bsy)$ as the eigenvalue of the operator $D=\sum_{a=1}^N x_a\p_a$,
\begin{equation}
D\psi_{\mathfrak{r}}(\bsx,\bsy)=|\mathfrak{r}|\psi_{\mathfrak{r}}(\bsx,\bsy),\quad\text{with}\quad |\mathfrak{r}|=\sum_{\alpha=1}^k |\bsr^{(\alpha)}|,\quad |\bsr^{(\alpha)}|=\sum_{a=1}^N \lfloor {r^{(\alpha)}_a}/{p} \rfloor=\sum_{a=1}^N n^{(\alpha)}_a,
\end{equation} 
where $\lfloor \bullet \rfloor$ is the floor function. The last equality is obtained from the correspondence $r_a^{(\a)}=pn_a^{(\a)}+i_a^{(\a)}-1$.

The functions $\psi_{\mathfrak{r}}(\bsx,\bsy)$ also satisfy the second condition of (\ref{eq:phys}), namely $J_{a,a}\psi_{\rf}(\bsx,\bsy)=k\psi_{\rf}(\bsx,\bsy)$, since each factor $\cY_{\boldsymbol{r}}(\bsx,\bsy)$ obeys the property
\begin{equation}
J_{a,a}\CY_\bsr(\bsx,\bsy)=\sum_{i=1}^p y_{i,a}\partial_{i,a}\cY_{\boldsymbol{r}}(\bsx,\bsy)=\cY_{\boldsymbol{r}}(\bsx,\bsy)
\end{equation}
for any $a$. In this sense, it gives an element of the physical space. We define the Hilbert space $\mathfrak{H}$ as the set of functions spanned by $\psi_{\mathfrak{r}}(\bsx,\bsy)$. It is important to note that the functions $\psi_{\mathfrak{r}}(\bsx,\bsy)$ for $\mathfrak{r}\in \mathfrak{R}$ are not linearly independent if $k\geq 2$, since the Pl\"ucker relations (\ref{Plucker}) produce nontrivial relations among the product of determinants.

The main result of this section is the general theorem below.
\begin{itembox}[l]{General form of the eigenstates}
The Hamiltonian action (\ref{Htilde_m}) is triangular on the set of functions $\psi_\mathfrak{r}(\bsx,\bsy)$ defined in (\ref{psil}),
\begin{equation}\label{trig}
	\tilde{\CH} \psi_{\mathfrak{r}}(\bsx,\bsy) =E(\mathfrak{r})\psi_{\mathfrak{r}}(\bsx,\bsy)+\sum_{|\mathfrak{s}|= |\mathfrak{r}|-2} C(\mathfrak{r},\mathfrak{s})\psi_{\mathfrak{s}}(\bsx,\bsy),
	\quad E(\mathfrak{r})=2|\mathfrak{r}|+N^2\,,
\end{equation}
where $C(\mathfrak{r},\mathfrak{s})$ is a numerical coefficient. As an immediate consequence of this formula, one obtains the eigenfunctions of the Hamiltonian through an appropriate recombination of the form,
\begin{equation}\label{general_eigenstate}
\Psi_{\mathfrak{r}}(\bsx,\bsy)=\psi_{\mathfrak{r}}(\bsx,\bsy)+\sum_{|\mathfrak{s}|\leq |\mathfrak{r}|-2} D(\mathfrak{r},\mathfrak{s})\psi_{\mathfrak{s}}(\bsx,\bsy)\,.
\end{equation}
These eigenfunctions have eigenvalue $E(\mathfrak{r})$.
\end{itembox}

The theorem (\ref{trig}) is one of the main results presented in this paper. Its proof is technical and relatively lengthy, it is given in appendix (\ref{sec:proof}). The second statement is a simple corollary obtained by a recursion procedure. Indeed, since the coefficients $C(\mathfrak{r},\mathfrak{s})$ are finite (as shown in the proof) and $E(\mathfrak{r})$ is a strictly positive integer, taking the combination 
\begin{equation}\label{subleading}
\Psi_{\mathfrak{r}}^{(1)}(\bsx,\bsy)=\psi_{\mathfrak{r}}(\bsx,\bsy)-\sum_{|\mathfrak{s}|=|\mathfrak{r}|-2}\frac{C(\mathfrak{r}, \mathfrak{s})}{E(\mathfrak{s})}\psi_{\mathfrak{s}}(\bsx,\bsy),
\end{equation}
we find that $\tilde{\CH}$ has a triangular action on $\Psi_{\mathfrak{r}}^{(1)}(\bsx,\bsy)$, and the r.h.s. now contains a linear combination of $\Psi_{\mathfrak{s}}^{(1)}$ with $|\mathfrak{s}|=|\mathfrak{r}|-4$. We can continue this recursive procedure until we arrive at the ground states, which are eigenstates since $|\rf|$ is minimal. While the functions $\psi_{\mathfrak{r}}(\bsx,\bsy)$ do not form a basis due to their redundancy, this fact is irrelevant here since we do not use independence. This set of functions provides a large class of solutions, and it is natural to conjecture that general eigenfunctions can be expressed as a linear combination of such functions. The equivalence relations among eigenstates, such as those following from the Pl\"ucker relations (\ref{Plucker}), will be discussed in section 5 below.

\paragraph{Ground state wave function and degeneracies} The ground states correspond to the labels $\rf$ for which $|\mathfrak{r}|$ is minimal. As a consequence of the triangular form \ref{trig} of the Hamiltonian action, the corresponding wave functions $\psi_{\rf}(\bsx,\bsy)$ are eigenfunctions of the system. We first examine the case $k=1$, and observe that the ground state is unique when $p$ divides $N$. It is obtained for $\bsr_0=(0,1,2,\cdots, N-1)$, and the corresponding wedge state is
\begin{equation}
\CY_{\bsr_0}(\bsx,\bsy)=[y_1\wedge\cdots\wedge y_p\wedge y_1x\wedge\cdots\wedge y_px\wedge\cdots\wedge y_1x^{m-1}\wedge\cdots\wedge y_px^{m-1}],\qquad (N=mp). 
\end{equation}
In the general case $N=mp+q$ with $0\leq q\leq p-1$, the ground state is $\left(\superp{p}{q}\right)$-fold degenerate which corresponds to a choice of spin for the $q$ extra particles,
\begin{equation}\label{vac_wedge}
\CY_{\bsr_0}(\bsx,\bsy)=[y_1\wedge\cdots\wedge y_p\wedge y_1x\wedge\cdots\wedge y_px\wedge\cdots\wedge y_1x^{m-1}\wedge\cdots\wedge y_px^{m-1}\wedge y_{i_1}x^m\wedge\cdots \wedge y_{i_q}x^m],\qquad (N=mp+q),
\end{equation} 
with $1\leq i_1<i_2<\cdots<i_q\leq p$. We denote $\mathfrak{R}_0$ the set of all possible ground states label, i.e. the set of labels $\bsr\in\mathfrak{R}$ that minimize $E(\bsr)$. The ground state energy for $N=mp+q$ is 
\begin{equation}
E(\bsr_0)=pm(m-1)+2qm+N^2.
\end{equation} 

When $k>1$, ground states wave functions are obtained by taking the product of $k$ determinants \ref{vac_wedge},
\begin{equation}\label{gnd_k}
\psi_{\rf_0}(\bsx,\bsy)=\D(\bsx)\prod_{\a=1}^{k}\CY_{\bsr_0^{(\a)}}(\bsx,\bsy),\quad E(\rf_0)=k\left(pm(m-1)+2qm\right)+N^2
\end{equation} 
with $\rf_0=(\bsr_0^{(1)},\cdots,\bsr_0^{(k)})$ and $\bsr_0^{(\a)}\in\mathfrak{R}_0$. Note that the expression of the ground state energy $E(\rf_0)$ reproduces the formula \ref{E0} obtained for the abelian model. It also agrees with the results obtained in \cite{Dorey2016}. The spin indices $i_l^{(\a)}$, $\a=1\cdots k$ and $l=1\cdots q$ labeling these states transform in the $k$-th fold symmetrization of the $q$th antisymmetric representation of $SU(p)$ \cite{Dorey2016}, and the dimension of $\mathfrak{R}_0$ is given by the hook length formula
\begin{equation}
\prod_{\alpha=1}^k\prod_{i=1}^q \frac{p+\alpha-i}{q+k+1-\alpha-i}
\end{equation} 
In fact, $\mathfrak{R}_0$ has more degrees of freedom than above because of the redundancies mentioned above.

Finally, we would like to make a comment on the excited states. The theorem (\ref{trig}) shows that excited states are linear combinations of products of the wedge states. They can be classified according to $U(p)$ (or $\gl(p)$) irreducible representations. In the $k=1$ case, it can be done using the Gelfand-Zetlin basis, this is explained in appendix \ref{app:GZ}. For general $k$, we have to introduce the corresponding loop (or Kac-Moody) algebra, and it will presented in the next section.

\subsection{KZ equation for the ground state wave function} 
The wave functions of the matrix model differs from those of the Hamiltonian by a Vandermonde factor,\footnote{We continue to omit the exponential factor $e^{-\frac{B}2\sum_a x_a^2}$ here.}
\begin{equation}\label{Phi}
\Phi_{\rf_0}(\bsx,\bsy)=\D(\bsx)^{-1}\psi_{\rf_0}(\bsx,\bsy)=\prod_{\a=1}^k \CY_{\bsr_0^{(\a)}}(\bsx,\bsy),
\end{equation} 
In appendix \ref{AppA}, we prove that these wave functions satisfy the following Knizhnik-Zamolodchikov (KZ) equation
\begin{equation}\label{eq:KZ}
(p+k)\partial_a \Phi_{\rf_0}(\bsx,\bsy)=\sum_{b\neq a}\frac{J_{a,b}J_{b,a}}{x_a-x_b}\Phi_{\rf_0}(\bsx,\bsy).
\end{equation}
This is a generalization of a similar result obtained in the case $N=mp$ in \cite{Dorey2016}. 
Indeed, in this case the ground state is non-degenerate, and the corresponding wave function $\Phi_{\rf_0}=(\CY_{\bsr_0})^k$ can be identified with the DTT wave function obtained as the conformal block function, $\langle V(x_1)\cdots V(x_N)\rangle$, where $V(x)$ is the primary field of the $k$-th order symmetric representation $[k]$, of a Wess-Zumino-Witten model $\widehat{\mathfrak{su}}(p)_k\otimes \widehat{\mathfrak{u}}(1)$. To identify the two expressions, we can re-label the particles and write the wedge state in the form
\begin{equation}
\CY_{\bsr_0}(\bsx,\bsy)=\e[y_1\wedge y_1 x\wedge\cdots \wedge y_1x^{m-1}\wedge y_2\wedge y_2 x\wedge\cdots\wedge y_2 x^{m-1}\wedge\cdots\cdots\wedge y_p\wedge y_p x\wedge \cdots\wedge y_p x^{m-1}],
\end{equation} 
where $\e$ is a sign coming from the permutation of the columns in the determinant. Expanding the determinant gives the antisymmetrization over all particles
\begin{equation}
\CY_{\bsr_0}(\bsx,\bsy)=\e\sum_{\s\in S_N}(-)^\s\prod_{a=1}^m \left[y_{1,\s(a)}(x_{\s(a)})^{a-1}\times y_{2,\s(a+m)}(x_{\s(a+m)})^{a-1}\times\cdots\times y_{p,\s(a+m(p-1))}(x_{\s(a+m(p-1))})^{a-1}\right]
\end{equation} 
this operation is denoted $\CA$ in \cite{Dorey2016}. It is then possible to anti-symmetrize independently the particles in each set of $m$ particles with the same spin. It produces the Vandermonde determinants found in the equation (5.5) of \cite{Dorey2016}, 
\begin{align}
\begin{split}
\CY_{\bsr_0}(\bsx,\bsy)&=\dfrac{\e}{(m!)^p}\sum_{\s\in S_N}(-)^\s\Bigg[\prod_{a=1}^m y_{1,\s(a)}\prod_{\superp{a,b=1}{a<b}}^{m}(x_{\s(a)}-x_{\s(b)})\times \prod_{a=m+1}^{2m}y_{2,\s(a)}\prod_{\superp{a,b=m+1}{a<b}}^{2m}(x_{\s(a)}-x_{\s(b)})\times\\
&\quad\quad\times\cdots\times \prod_{a=m(p-1)+1}^{mp}y_{p,\s(a)}\prod_{\superp{a,b=(p-1)m+1}{a<b}}^{pm}(x_{\s(a)}-x_{\s(b)})\Bigg]\\
&=\dfrac{\e}{(m!)^p}\CA\Bigg[\prod_{a=1}^m y_{1,a}\prod_{\superp{a,b=1}{a<b}}^{m}(x_a-x_b)\times \prod_{a=m+1}^{2m}y_{2,a}\prod_{\superp{a,b=m+1}{a<b}}^{2m}(x_{a}-x_{b})\times\cdots\times \prod_{a=m(p-1)+1}^{mp}y_{p,a}\prod_{\superp{a,b=(p-1)m+1}{a<b}}^{pm}(x_{a}-x_{b})\Bigg].
\end{split}
\end{align}

In the next section, we show that the $U(p)$ Kac-Moody symmetry appears in the large $N$-limit. We relate the KZ equation (\ref{eq:KZ}) to the correlation function of the vertex operators, which intertwine the wedge states with the states of the free fermion Fock space. The results of this subsection show that the KZ equation holds even at finite $N$.

We note that the Calogero model ground states give a limited class of the chiral correlators of the WZW model. For the study of more general correlators in the context of the FQHE, see for example, \cite{ardonne2010chiral}\cite{fukusumi2023operator}.

\subsection{Generalized Statistics}
Let us focus for a moment on the vacuum wave functions in the case $k=1$ for which the matrix model wave functions coincide with wedge states, i.e. $\Phi_{\bsr_0}(\bsx,\bsy)=\CY_{\boldsymbol{r_0}}(\bsx,\bsy)$. The latter can be written in the explicit form
\begin{equation}
    \cY_{\bsr_0}(\bsx,\bsy)=\left|
    \begin{array}{ccccccc}
         y_{1,1}& \cdots & y_{p,1}& \cdots & y_{i_1,1}x_1^{m}& \cdots &  y_{i_q,1}x_1^{m}\\
         y_{1,2}& \cdots & y_{p,2}& \cdots & y_{i_1,2}x_2^{m}& \cdots &  y_{i_q,2}x_2^{m}\\
         \vdots & & & \ddots & & &  \vdots\\
         y_{1,N} &\cdots & y_{p,N} & \cdots & y_{i_1,N}x_N^m&\cdots & y_{i_q,N}x_N^m
    \end{array}
    \right|\,.
\end{equation}
An important feature of the wave functions $\CY_{\bsr_0}(\bsx,\bsy)$ is that they do not vanish when two coordinates coincide $x_a\to x_b$ because of the presence of the extra spin variables $y_{i,a}$. Instead, when $p+1$ particles approach each other at the same time, at least two particles will have the same spin. As a result, the two corresponding rows of the determinants will be identical, and the determinant will vanish. More precisely, when $N=mp+q$, all the wave functions $\CY_{\bsr_0}(\bsx,\bsy)$ vanish when $p+1$ particles coordinate coincide. When only $r<p+1$ particles coincide, some of the wave functions might still vanish if $q>0$, which partially lifts the degeneracy of the vacuum. In this way, the wedge states describe particles obeying a generalized exclusion principle.


\paragraph{Fractional quasi-holes} Using a similar argument, it is possible to introduce wave functions describing fractional quasi-holes,
\begin{equation}
    \cY_{\bsr_0}(\bsx,\bsy|\bszeta)=\left|
    \begin{array}{ccccccc}
         y_{1,1}& \cdots & y_{p,1}& \cdots & y_{i_1,1}x_1^{m+1}& \cdots &  y_{i_q,1}x_1^{m+1}\\
         \vdots & & & \ddots & & &  \vdots\\
         y_{1,N} &\cdots & y_{p,N} & \cdots & y_{i_1,N}x_N^{m+1}&\cdots & y_{i_q,N}x_N^{m+1}\\
         y_{1,N+1} &\cdots & y_{p,N+1} & \cdots & y_{i_1,N+1}\zeta_1^{m+1}&\cdots & y_{i_q,N+1}\zeta_1^{m+1}\\
         \vdots & & & \ddots & & &  \vdots\\
         y_{1,N+p} &\cdots & y_{p,N+p} & \cdots & y_{i_1,N+p}\zeta_p^{m+1}&\cdots & y_{i_q,N+p}\zeta_p^m
    \end{array}
    \right|
\end{equation}
When we set $\zeta_1=\cdots=\zeta_p=\zeta$, the wave functions $\cY_{\bsr_0}(\bsx,\bsy|\bszeta)$ contain a factor $\prod_{a=1}^N (\zeta-x_a)$ due to the generalized exclusion principle mentioned above, and it describes a hole of coordinate $\zeta$. Thus, the general formula $\cY_{\bsr_0}(\bsx,\bsy|\bszeta)$ describes the splitting of this hole into $p$ quasi-holes, each of which has a fractional charge $-e/p$.

\paragraph{Moore-Read state} Some of the characteristic states of the non-Abelian FQHE can also be recovered from our general formula for the ground states wave function. For instance, taking $p=2$, $N=2m$, and $k=2$, we find a singlet ground state with parameters $\boldsymbol{r}_0=\{0,1,\cdots, 2m-1\}$. The corresponding matrix model wave function is
\begin{equation}
    \Phi(\bsx,\bsy)=(\cY_{\boldsymbol{r}_0}(\bsx,\bsy))^2\,.
\end{equation}
It was noted in \cite{Dorey2016} that this expression is equivalent to the following Pfaffian,\footnote{To recover our expression, we need to substitute  $z_a\to x_a$ and $\zeta_a\to y_{1,a}/y_{2,a}$ in the equation (3.7) of \cite{Dorey2016}.}
\begin{equation}\label{Moore-Read}
    \Phi(\bsx,\bsy)\propto
    \mbox{Pf}\left(
    \frac{(y_{1a}y_{2b}-y_{2a}y_{1b})^2}{x_a-x_b}
    \right)\prod_{a<b} (x_a-x_b)\,,
\end{equation}
The coordinate dependence of this wave function coincides with the Moore-Read state.

It seems possible to generalize the Pfaffian formula for general $p=k$ and $N=pm$ with $\boldsymbol{r}_0=\{0,1,\cdots, pm-1\}$ in the following way
\begin{equation}
    \Phi(\bsx,\bsy)=(\cY_{\boldsymbol{r}_0}(\bsx,\bsy))^p
    \propto \D(\bsx)  \sum_{A^1,\cdots, A^m}\epsilon(A^1,\cdots, A^m) \prod_{\alpha=1}^m\frac{(\det_{i,j}y_{i,a^\alpha_j})^p}{\prod_{1\leq i<j\leq p} (x_{a^\alpha_i}-x_{a^\beta_j})}
\end{equation}
The summation is defined over all possible decomposition of $\left\{1,2,\cdots, pm\right\}$ into $L$ sets $A^\alpha$ of $p$ mutually distinct indices. Namely, $A^{\alpha}=\left\{a^\alpha_1,\cdots, a^\alpha_p\right\}$ ($a^\alpha_i> a^\alpha_j$ for $i>j$), and $a^\alpha_i\in \left\{1,2,\cdots, pm\right\}$, $A^\alpha\cap A^\beta =\emptyset$ and $\bigcup_{\alpha=1}^p A^\alpha=\left\{1,2,\cdots, pm\right\}$. The sign $\epsilon(A^1,\cdots, A^m)$ depends on the sets $A^\alpha$. This conjecture is a natural generalization of the Moore-Read wave function (\ref{Moore-Read}). We have checked that the power counting for both $x$ and $y$ variables match separately, and observed the correct behavior for the generalized statistics. Indeed, the wave function vanishes only when $p+1$ coordinates $x_a$ approach each other, and such behavior is broken when the $y$ vectors become linearly dependent.




\section{Free fermion representation and Kac-Moody algebra}
In the previous section, we have shown that the eigenstates of the Hamiltonian can be expressed as linear combinations of products of $k$ wedge states. However, the Pl\"ucker identities imply that this description is redundant for $k>1$. In this section, we rewrite the eigenstates using a free fermion formalism \cite{DKJM1983}. This has two main advantages. First, it allows us to re-interpret the Pl\"ucker relations at finite $N$ as constraints from a loop-algebra symmetry. Second, it brings to light the appearance of the Kac-Moody symmetry $\widehat{\mathfrak{su}}(p)_k$ in the large $N$-limit. 

\subsection{Pl\"ucker identity as \texorpdfstring{$\widehat{\mathfrak{su}}(k)_+$}{su(k)}-constraints at finite \texorpdfstring{$N$}{N}}
For a finite number $N$ of particles, we introduce a set of (non-relativistic) free fermion oscillators, $\psi^{i,\alpha}_n$, $\bar\psi^{i,\alpha}_n$, indexed by the integers $n\geq 0$, $i=1,\cdots, p$, $\alpha=1,\cdots, k$, and satisfying
\begin{equation}
    \left\{\psi^{i,\alpha}_n, \bar\psi^{j,\beta}_m\right\}=\delta_{i,j}\delta_{\alpha,\beta}\delta_{n,m},\quad
    \left\{\psi^{i,\alpha}_n, \psi^{j,\beta}_m\right\}=\left\{\bar\psi^{i,\alpha}_n, \bar\psi^{j,\beta}_m\right\}=0\,.
\end{equation}
We also define the bra and ket vacua as,
\begin{equation}\label{fermions}
    \psi^{i,\alpha}_n|0\rangle=0,\quad
    \langle 0|\bar\psi^{i,\alpha}_n=0\,,\quad \mbox{ for }n\geq 0\,.
\end{equation}
Generic states of the free fermion Fock space are obtained by the action of the modes $\bar\psi_n^{i,\a}$ on the vacuum $\ket{0}$. This Fock space has a natural family of grading operators $\hat N_\alpha=\sum_{n=0}^\infty\sum_{i=1}^p \bar\psi^{i,\alpha}_n \psi^{i,\alpha}_n$ corresponding to the fermionic numbers counting the number of modes $\bar\psi_{n}^{i,\alpha}$ for each $\a=1\cdots k$. As a result, the total Fock space $\CF^{(k)}$ decomposes as a direct sum of subspaces $\mathcal{F}^{(k)}_{\boldsymbol{N}}$ labeled by the vector $\boldsymbol{N}=(N_1,\cdots, N_k)$ encoding the eigenvalues $N_\a\in\mZ^{\geq0}$ of the fermionic number operators $\hat N_\alpha$. These subspaces admit the following basis
\begin{equation}
    \left\{\prod_{\alpha=1}^k\left( \bar\psi^{i_1^{(\alpha)},\alpha}_{n_1^{(\alpha)}}\cdots \bar\psi^{i_{N_\alpha}^{(\alpha)},\alpha}_{n_{N_\alpha}^{(\alpha)}}\right)|0\rangle,\quad n_{a_\a}^{(\a)}\in\mZ^{\geq0},\ i_{a_\a}^{(\a)}\in[\![1,p]\!],\quad a_\a=1\cdots N_\a\right\} 
\end{equation}
In the following, we restrict ourselves to the case $\boldsymbol{N}_0=(N,\cdots, N)$ where $N$ is the number of particles, and denote the corresponding subspace $\CF_{\boldsymbol{N}_0}^{(k)}$.

Let $\mathcal{V}^{(k)}_{N}$ denote the space linearly generated by the wave functions of the matrix model (i.e., omitting the Vandermonde factor in $\psi_{\rf}(\bsx,\bsy)$). As shown in the previous section, this space is spanned by products of $k$ wedge states (which are not all linearly independent),
\begin{equation}\label{sols}
    \mathcal{V}^{(k)}_{N}=\Span_{(\bsr^{(1)},\cdots,\bsr^{(k)})\in\mathfrak{R}}\left\{\prod_{\alpha=1}^k \cY_{\bsr^{(\a)}}(\bsx,\bsy)\right\}.
\end{equation}
We define a map from the fermionic Fock space $\mathcal{F}^{(k)}_{\boldsymbol{N}_0}$ to the wave functions $\mathcal{V}^{(k)}_{N}$ realized as a linear map\footnote{The second identity follows from the anti-commutation relation $    \left\{\psi^\alpha(x_a,y_a), \bar\psi^{i,\beta}_n\right\}=\delta_{\alpha\beta} x_a^n y_{i,a}$.}
\begin{align}
\begin{split}
&    \prod_{\alpha=1}^k\left( \bar\psi^{i_1^{(\alpha)},\alpha}_{n_1^{(\alpha)}}\cdots \bar\psi^{i_N^{(\alpha)},\alpha}_{n_N^{(\alpha)}}\right)|0\rangle\to
    \langle \boldsymbol{x},\boldsymbol{y}|
    \prod_{\alpha=1}^k\left( \bar\psi^{i_1^{(\alpha)},\alpha}_{n_1^{(\alpha)}}\cdots \bar\psi^{i_N^{(\alpha)},\alpha}_{n_N^{(\alpha)}}\right)|0\rangle=\e\prod_{\alpha=1}^k [x^{n^{(\alpha)}_{1}}y_{i^{(\alpha)}_{1}}\wedge \cdots \wedge x^{n^{(\alpha)}_{N}}y_{i^{(\alpha)}_{N}}]
     \,,
\end{split}
\end{align}
using the bra state $\langle \boldsymbol{x}, \boldsymbol{y}|$ defined as\footnote{Note that the free fermion operator $\psi^\alpha(x,y)$ can also be written as $\psi^\alpha(z)=\sum_{n=0}^\infty\sum_{i=1}^p \psi^{i,\alpha}_n z^{pn+i}$ using the translation rule given in section 3.}
\begin{align}
     \langle \boldsymbol{x}, \boldsymbol{y}| &=\langle 0|\Psi(x_N,y_{N})\cdots\Psi(x_1,y_1)\\
    \Psi(x_a, y_a)&=\prod_{\alpha=1}^k\psi^\alpha (x_a,y_a) ,\quad \psi^\alpha (x_a,y_a)=\sum_{n=0}^\infty \sum_{i=1}^p \psi^{i,\alpha}_n x_a^n y_{i,a}\,.
\end{align}
Here $\e=\pm$ is a sign that will be irrelevant in our discussion. It is important to note that the fermionic states \ref{fermions} are linearly independent and define a basis of the space $\mathcal{F}^{(k)}_{\boldsymbol{N}_0}$. However, we have seen that this is not the case for the products of wedge states obtained from the inner product with $\langle \boldsymbol{x}, \boldsymbol{y}|$. We will now show that the linear dependence is a consequence of the invariance of the bra state under a loop algebra that we denote $\widehat{\mathfrak{su}}(k)_+$.

\paragraph{$\widehat{\mathfrak{su}}(k)_+$-symmetry} Let $L_n^{\a\b}$ with $n\geq0$ and $\a,\b\in[\![1,k]\!]$ denote the positive generators of the loop algebra of $\mathfrak{u}(k)$ in the fermionic representation,
\begin{equation}\label{gln-loop}
    L^{\alpha\beta}_n = \sum_{\ell=0}^\infty \sum_{i=1}^p \bar\psi^{i,\a}_{n+\ell}\psi^{i,\b}_{\ell}\,,
    \qquad (n\geq 0).
\end{equation}
We note that the fermionic number operators defined earlier correspond to $\hat N_\a=L_0^{\a\a}$. These generators satisfy the commutation relations
\begin{equation}
    \left[L^{\alpha\beta}_n, L^{\gamma\delta}_m\right]=
    \delta_{\beta,\gamma}L^{\alpha\delta}_{n+m}-\delta_{\alpha,\delta}L^{\gamma\beta}_{n+m}\,,
\end{equation}
and we denote $\widehat{\mathfrak{u}}(k)_+$ the algebra spanned by these operators. These operator act on the free fermion Fock spaces $\CF^{(k)}$, and obey the following commutation relations with the fermionic fields,
\begin{equation}
[L^{\alpha\beta}_n,\psi^\gamma(x_a,y_a)]=-\delta_{\alpha,\gamma}x_a^{n}\psi^\beta(x_a,y_a),\quad
[L^{\alpha\beta}_n,\Psi(x_a,y_a)]=-\delta_{\alpha,\beta}x_a^{n}\Psi(x_a,y_a)\,,
\end{equation}
where the second identity is a consequence of the fermionic exclusion principle $(\psi^{\alpha}(x,y))^2=0$. We deduce the action of $\widehat{\mathfrak{u}}(k)_+$ generators on the bra states $\bra{\bsx,\bsy}$ introduced earlier,
\begin{equation}\label{glk-constraint}
    \langle \boldsymbol{x}, \boldsymbol{y}| L^{\alpha\beta}_n=\delta_{\alpha,\beta}p_n(\bsx)
   \langle \boldsymbol{x}, \boldsymbol{y}|\,,
\end{equation}
where $p_n(\bsx)=\sum_{a=1}^N x_a^{n}$ is the elementary power sum symmetric polynomials. It implies that the bra state is annihilated by the action of the $\widehat{\mathfrak{su}}(k)_+$ generators defined as,
\begin{equation}\label{slk-constraint}
    \langle \boldsymbol{x}, \boldsymbol{y}|\bar{L}^{\alpha\beta}_n=0\,,\qquad
    \bar{L}^{\alpha\beta}_n = L^{\alpha\beta}_n-\frac{\delta_{\alpha,\beta} }{k}\sum_{\gamma} L^{\gamma\gamma}_n.
\end{equation}
It means that the map from the fermionic Fock space $\CF^{(k)}_{\boldsymbol{N}_0}$ to the space of wedge states products $\CV_N^{(k)}$ is invariant under the $\widehat{\mathfrak{su}}(k)_+$ symmetry. This invariance explains the redundancy of our description of $\mathcal{V}^{(k)}_N$. In the following, we refer to the subspace of $\mathcal{F}^{(k)}_{\boldsymbol{N}}$ obtained from the action of the generators $\bar{L}^{\alpha\beta}_n$ on the fermionic vacuum as the $\widehat{\mathfrak{su}}(k)_+$-orbit. As we have shown, any state from this orbit has a vanishing inner product with $\bra{\bsx,\bsy}$, and thus belongs to the kernel of the map $\CF^{(k)}_{\boldsymbol{N}_0}\to\CV_N^{(k)}$.

The Pl\"ucker relations (\ref{Plucker}) are a consequence of the $\widehat{\mathfrak{su}}(k)_+$-symmetry. To show this, we restrict ourselves to $k=2$ and consider the action of the generator $\bar L^{21}_n=L_n^{21}$ on the Fock space $\CF_{\boldsymbol{N}}^{(k)}$ with $\boldsymbol{N}=(N+1,N-1)$. The action on the basis vectors reads 
\begin{align*}
    &L^{21}_n \bar\psi^{i_1,1}_{n_1}\cdots\bar\psi^{i_{N+1},1}_{n_{N+1}}
    \bar\psi^{j_1,2}_{m_1}\cdots\bar\psi^{j_{N-1},2}_{m_{N-1}}|0\rangle\\
    &~~~~~~~~~=(-1)^N\sum_{a=1}^{N+1}(-1)^{a-1}
\bar\psi^{i_1,1}_{n_1}\cdots\,\cancel{\bar\psi^{i_a,1}_{n_a}}\,\cdots \bar\psi^{i_{N+1},1}_{n_{N+1}}  \bar\psi^{i_a,2}_{n_a+n}\bar\psi^{j_1,2}_{m_1}\cdots\bar\psi^{j_{N-1},2}_{m_{N-1}}|0\rangle
\end{align*}
After taking the inner product with $\langle \boldsymbol{x},\boldsymbol{y}|$ and using (\ref{slk-constraint}), we find the identity
\begin{equation}\label{generalized-Plucker}
    \sum_{a=1}^{N+1}(-1)^{a} [y_{i_1}x^{n_1}\wedge \cdots \wedge\, \cancel{y_{i_a}x^{n_a}}\wedge \cdots \wedge y_{i_{N+1}}x^{n_{N+1}}]\cdot[y_{i_a}x^{n_a+n}\wedge y_{j_1}x^{m_1}\wedge \cdots \wedge y_{j_{N-1}}x^{m_{N-1}}]=0.
\end{equation}
This is indeed a generalization of the Pl\"ucker relations (\ref{Plucker}) which are recovered by taking $n=0$. The same relations are obtained if consider instead the action of $\bar L^{12}_n$ on $\CF_{\boldsymbol{N}}^{(k)}$ with $\boldsymbol{N}=(N-1,N+1)$.

It is also instructive to examine the action of the diagonal element $\bar L_n^{11}=\frac12(L_n^{11}-L_n^{22})=-\bar L_n^{22}$ on the basis vectors of the Fock space $\CF_{\boldsymbol{N}_0}^{(k)}$,
\begin{align*}
   & (L^{11}_n-L^{22}_n)\bar\psi^{i_1,1}_{n_1}\cdots\bar\psi^{i_{N},1}_{n_{N}}
\bar\psi^{j_1,2}_{m_1}\cdots\bar\psi^{j_{N},2}_{m_{N}}|0\rangle\\
 &~~~~~~=\sum_{a=1}^N\left(
\bar\psi^{i_1,1}_{n_1}\cdots\,\bar\psi^{i_a,1}_{n_a+n}\,\cdots\bar\psi^{i_{N},1}_{n_{N}}
\bar\psi^{j_1,2}_{m_1}\cdots\bar\psi^{j_{N},2}_{m_{N}}-
\bar\psi^{i_1,1}_{n_1}\cdots\bar\psi^{i_{N},1}_{n_{N}}
\bar\psi^{j_1,2}_{m_1}\cdots\bar\psi^{j_a,2}_{m_a+n}\,\cdots\bar\psi^{j_{N},2}_{m_{N}}
 \right)|0\rangle\,.
\end{align*}
After taking the inner product with the bra state $\bra{\bsx,\bsy}$, we find an identity which is trivially satisfied,
\begin{align*}
&    \sum_{a=1}^N\left( [y_{i_1}x^{n_1}\wedge \cdots \wedge y_{i_a}x^{n_a+n}\wedge \cdots \wedge y_{i_N}x^{n_N}]\cdot [y_{j_1}x^{m_1}\wedge \cdots \wedge y_{j_N}x^{m_N}]\right.\\
&~~~~~-\left.
    [y_{i_1}x^{n_1}\wedge \cdots \wedge y_{i_N}x^{n_N}]\cdot[y_{j_1}x^{m_1}\wedge \cdots \wedge y_{j_a}x^{m_a+n}\wedge \cdots \wedge y_{j_N}x^{m_N}]\right)\\
&=(p_n(\bsx)[y_{i_1}x^{n_1}\wedge \cdots \wedge y_{i_N}x^{n_N}])[y_{j_1}x^{m_1}\wedge \cdots \wedge y_{j_N}x^{m_N}]\\
&~~~~~~~~-[y_{i_1}x^{n_1}\wedge \cdots \wedge y_{i_N}x^{n_N}](p_n(\bsx)[y_{j_1}x^{m_1}\wedge \cdots \wedge y_{j_N}x^{m_N}])\\
&=0\,.
\end{align*}
This shows that the invariance of the bra state $\bra{\bsx,\bsy}$ under $\widehat{\mathfrak{su}}(k)_+$-symmetry for $k=2$ is equivalent to the generalized Pl\"ucker identities (\ref{generalized-Plucker}) obeyed by products of wedge states.


\paragraph{Ground states} The ground states wave functions (\ref{gnd_k}) for $N=pm+q$ (with $0\leq q <p$) are associated to the following states of $\mathcal{F}^{(k)}_{\boldsymbol{N}_0}$,
\begin{equation}\label{gnd_fermion}
    \prod_{\alpha=1}^k \prod_{r=1}^q \bar\psi^{i_r^{(\alpha)},\alpha}_{m}
    \prod_{\alpha=1}^k\prod_{i=1}^p\prod_{l=1}^{m-1} \bar\psi^{i,\alpha}_l |0\rangle
\end{equation}
When $q=0$, the first factor is absent, and the vacuum is unique. When $q>0$, the vacua are labeled by the indices $(i_1^{(\a)},\cdots,i_q^{(\a)})$ with $\a=1\cdots k$ taking value in $[\![1,p]\!]$. The vacua form a representation $[k^q]$ of $\mathfrak{su}(p)$ \cite{Dorey2016}. They belong to a dual $[q^k]$ representation of $\mathfrak{su}(k)$, which is singlet, which does not belong to the $\widehat{\mathfrak{su}}(k)_+$-orbit.\footnote{We note that the orbit cannot contain a singlet state since $L^{\alpha\beta}_n|\chi\rangle\neq 0$ implies that neither $|\chi\rangle$ nor $L^{\alpha\beta}_n|\chi\rangle$ belong to the singlet.}

\paragraph{Dual loop symmetry and excited states} It is possible to define an action of the larger algebra $\widehat{\mathfrak{u}}(pk)_+$ on the fermionic Fock space $\CF^{(k)}$. This algebra is defined as the positive part of the loop algebra of $\mathfrak{u}(pk)$, it is generated by the following operators in the fermionic representation
\begin{equation}
        M^{i+p(\alpha-1),j+p(\beta-1)}_n = \sum_{\ell=0}^\infty  \bar\psi^{i,\alpha}_{n+\ell}\psi^{j,\beta}_{\ell}\,,
    \qquad (n\geq 0)
\end{equation}
where the indices $i+p(\alpha-1)$ and $j+p(\beta-1)$ run from $1$ to $pk$. 

The algebra $\widehat{\mathfrak{u}}(pk)_+$ contains a subalgebra $\widehat{\mathfrak{u}}(p)_+$ generated by
\begin{equation}
    K^{ij}_n = \sum_{\ell=0}^\infty  \sum_{\alpha=1}^k \bar\psi^{i,\alpha}_{n+\ell}\psi^{j,\alpha}_{n}\,,
    \qquad (n\geq 0),
\end{equation}
which is dual to the algebra $\widehat{\mathfrak{u}}(p)_+$ in the sense that
\begin{equation}
    \left[L^{\alpha\beta}_n, K^{ij}_m\right]=0\,.
\end{equation}
More precisely, we have the decomposition \cite{Nakanishi:1990hj}
\begin{equation}\label{decomp1}
    \widehat{\mathfrak{u}}(pk)_+\supset \widehat{\mathfrak{su}}(p)_+\oplus \widehat{\mathfrak{su}}(k)_+\oplus \widehat{\mathfrak{u}}(1)_+
\end{equation}
where the $\widehat{\mathfrak{u}}(1)_+$ subalgebra is generated by
\begin{equation}
J_n=\sum_{\ell=0}^\infty  \sum_{\alpha=1}^k\sum_{i=1}^p \bar\psi^{i,\alpha}_{n+\ell}\psi^{i,\alpha}_{n}=\sum_{\a=1}^k L_n^{\a\a}=\sum_{i=1}^p K_n^{ii}.
\end{equation}
It is obvious that the states generated by the action of $\widehat{\mathfrak{su}}(p)_+\oplus\widehat{\mathfrak{u}}(1)_+$ on the vacuum
do not belong to the $\widehat{\mathfrak{su}}(k)_+$-orbit. In the large $N$ limit, it is possible to show that these states exhaust the possible eigenstates of the Hamiltonian \ref{Htilde_m}, and we expect that this statement holds also at finite $N$.

The adjoint action of the $\widehat{\mathfrak{u}}(p)_+$ generators on the fermionic field $\Psi(x_a,y_a)$ reads
\begin{equation}\label{KMprimary}
    \left[K^{ij}_n,\Psi(x_a,y_a)\right]=-x_a^n y_{i,a}\partial_{j,a}\Psi(x_a,y_a),
\end{equation}
which shows that the fields $\Psi(x,y)$ is the finite $N$ equivalent of the primary fields for the Kac-Moody algebra $\widehat{\mathfrak{u}}(p)$. The KZ equation for the ground state wave functions obtained in (\ref{eq:KZ}) is expected to follow from this primary field condition. Indeed, at large $N$ the derivation of the KZ equation from the primary field condition is a standard procedure \cite{knizhnik1984current}. On the other hand, the derivation of the KZ equation at finite $N$ given in the appendix \ref{A:KP} does not involve the $\widehat{\mathfrak{u}}(p)_+$-invariance, and it would be interesting to provide an alternative derivation starting from the condition (\ref{KMprimary}).

In the Abelian case ($p=1$), the algebraic structure simplifies and the Fock space $\CF^{(k)}$ is generated by the action of the $\widehat{\mathfrak{u}}(1)_+$ generators on the vacuum. Thus, in the large $N$ limit, the spectrum of the Hamiltonian (\ref{Htilde_m}) produces the $\widehat{\mathfrak{u}}(1)$ character \ref{U(1)character}, in agreement with the results obtained by Polychronakos in \cite{Polychronakos2001}.

\subsection{Kac-Moody algebra at large \texorpdfstring{$N$}{N} and level-rank duality}
In this subsection, we study the large $N$ limit of the fermionic Fock space $\CF^{(k)}$ and observe the emergence of the Kac-Moody symmetry $\widehat{\mathfrak{su}}(p)_k$. The emergence of this symmetry is one of the main results obtained by DTT in \cite{Dorey2016, Dorey:2016hoj}, it was later proved rigorously in \cite{Hu:2023eyx}. Our spin Calogero model provides an alternative point on view which we explain here.

In order to take the large $N$ (or, equivalently, large $m$) limit of the wedge states, we focus first on the vacuum wave functions (\ref{Phi}, \ref{vac_wedge})
\begin{align}
     \Phi_{\rf_0}(\bsx,\bsy)&=\prod_{\alpha=1}^k \cY_{\boldsymbol{r_0^{(\alpha)}}}(\bsx,\bsy)\label{Phi1}\\
   \cY_{\boldsymbol{r}_0^{(\alpha)}}(\bsx,\bsy)&=\left(\prod_{a=1}^N x_a^{m-1}\right) [y_{1}x^{-m+1}\wedge \cdots \wedge y_{p}x^{-m+1}\wedge \cdots\wedge
    y_1 \wedge \cdots\wedge y_p \wedge x y_{i_1^{(\alpha)}}\wedge \cdots \wedge x y_{i_q^{(\alpha)}}]\,.\label{Yralpha1}
\end{align}
with the decomposition $N=pm +q$, $0\leq q<p$. The factor $\prod_{a=1}^N x_a^{m-1}$ can be absorbed by a constant shift of the Hamiltonian, it can be ignored for the purpose of taking the large $N$ limit. Permuting the factors in the wedge product, we obtain an expression of the wedge states appropriate for the limit $N\to\infty$, it produces
\begin{equation}
    \cY_{\boldsymbol{r}_0^{(\alpha)}}^\infty=[(y_{i_1^{(\alpha)}}x)\wedge\cdots \wedge (y_{i_q^{(\alpha)}}x)\wedge y_1 \wedge \cdots\wedge y_p \wedge y_1 x^{-1}\wedge \cdots\wedge y_p x^{-1} \wedge \cdots ]\,.
\end{equation}
When $q=0$, the factors $y_{i_1^{(\alpha)}}x$ are not present, and we recover the usual free fermion vacuum written as a semi-infinite wedge product \cite{Uglov:1996np}. We will explain below how to introduce the extra factors when $q\neq 0$. General wedge states are obtained by replacing a finite number of factors $y_i x^{-n}$ by $y_i x^{-n+r}$ for some $r\in\mZ^{>0}$. The limiting procedure is the same for these states. We denote the space of the eigenfunctions $\Phi_{\rf}^{\infty}(\bsx,\bsy)$ built upon semi-infinite wedge products as $\mathcal{V}_\infty^{(k)}$. We note that, in the large $N$ limit, the wedge states become true eigenstates of the Hamiltonian (\ref{Htilde_m}) since the energies $E(\boldsymbol{\mathfrak{r}})$ grow as $O(N^2)$, and so the subleading terms in (\ref{subleading}) vanish.
We note that the coefficients $C(\boldsymbol{\mathfrak{r}},\boldsymbol{\mathfrak{s}})$ are finite, as seen from the explicit computation in Appendix A, and the number of the sets $\boldsymbol{\mathfrak{s}}$ is also finite.

To describe the Fock space at large $N$, we introduce the free fermionic modes $\psi^{i,\alpha}_r$, $\bar\psi^{i,\alpha}_r$, which still carry the color index $i=1,\cdots, p$ and Chern-Simons level index $\alpha=1,\cdots, k$, but now have the mode index $r\in \mathbb{Z}+\frac12$. These operators satisfy the anticommutation relations
\begin{equation}
    \left\{\psi^{i,\alpha}_r, \bar\psi^{j,\beta}_s\right\}=\delta_{i,j}\delta_{\alpha,\beta}\delta_{r+s,0},\quad
    \left\{\psi^{i,\alpha}_r, \psi^{j,\beta}_s\right\}=\left\{\bar\psi^{i,\alpha}_r, \bar\psi^{j,\beta}_s\right\}=0\,.
\end{equation}
The Fock space is built on a vacuum $\ket{\vac}$ which can be seen formally as the limit of the ground state (\ref{gnd_fermion}) for $q=0$.
The action of the fermionic modes on this vacuum and the corresponding dual state reads
\begin{equation}
\psi^{i,\alpha}_r|\vac\rangle=\bar\psi^{i,\alpha}_r|\vac\rangle=0,\quad
    \langle \vac|\psi^{i,\alpha}_{-r}=\langle \vac|\bar\psi^{i,\alpha}_{-r}=0\,,\qquad (r>0)
\end{equation}
We denote $\CF_\infty^{(p,k)}$ the Fock space obtained by the action of negative modes $\psi^{i,\alpha}_{-r}$, $\bar\psi^{i,\alpha}_{-r}$ on this vacuum.

The map $\CF_{\boldsymbol{N}_0}^{(k)}\to\CV_N^{(k)}$ defined in the previous subsection can be extended formally to the limit of infinite number of particles, i.e. $N=mp+q$ with $m\to\infty$, $q\in[\![0,p-1]\!]$ fixed. To do so, we extend the definition of the bra state $\bra{\bsx,\bsy}$ as follows,
\begin{align}
    \langle \boldsymbol{x}, \boldsymbol{y}|&= \lim_{m\to \infty} \langle\vac| :\left(\prod_{\alpha=1}^k \prod_{i=1}^p \prod_{r=1/2}^{m-1/2}\bar\psi^{i,\alpha}_{r}\right)\Psi(x_{mp+q},y_{mp+q})\cdots \Psi(x_1,y_1):\\
    \Psi(x_a,y_a)& =\prod_{\alpha=1}^k \psi^\alpha(x_a,y_a),\quad
    \psi^\alpha(x_a,y_a) =\sum_{r\in \mathbb{Z}+1/2} \sum_{i=1}^p \psi^{i,\alpha}_r x_a^{r+1/2}y_{i,a}
\end{align}
where $:\cdots:$ is the fermionic normal ordering. We use the anti-commutation rule,
\begin{equation}
    \left\{\psi^\alpha(x,y), \bar\psi^{i,\beta}_r\right\}=x^{-r+1/2} y_i \,,
\end{equation}
to show that
it defines by projection a map $\CF_\infty^{(p,k)}\to\CV_\infty^{(k)}$ which sends the vacuum state $\ket{\vac}$ to the ground state wave function (\ref{Phi1}) for $q=0$. The ground states wave functions for $q\neq 0$ can be obtained from the fermionic states 
\begin{equation}
\prod_{\alpha=1}^k \prod_{r=1}^q \bar\psi^{i_r^{(\alpha)},\alpha}_{-1/2}|\vac\rangle\,.
\end{equation}

\paragraph{$\widehat{\mathfrak{u}}(pk)$-symmetry and level-rank duality} The fermionic Fock space $\CF_\infty^{(p,k)}$ admits the action of the $\widehat{\mathfrak{u}}(pk)_1$ Kac-Moody algebra generated by the operators
\begin{equation}
    J^{i\alpha;j\beta}_n =\sum_{s\in\mZ+1/2} :\bar\psi^{i\alpha}_s \psi^{j,\beta}_{n-s}:\,,
\end{equation}
The indices $i=1\cdots p$, and $\alpha=1\cdots k$ can be combined into a single index $i+p(\alpha-1)$ running from one to $pk$. This algebra contains two commuting Kac-Moody subalgebras $\widehat{\mathfrak{u}}(p)_k$ and  $\widehat{\mathfrak{u}}(k)_p$ respectively generated by
\begin{equation}
    K^{ij}_n  = \sum_{s\in\mZ+1/2} \sum_{\alpha=1}^k :\bar\psi^{i\alpha}_s \psi^{j,\alpha}_{n-s}:\,,\quad
    L^{\alpha\beta}_n  = \sum_{s\in\mZ+1/2} \sum_{i=1}^p :\bar\psi^{i\alpha}_s \psi^{i,\beta}_{n-s}:.
\end{equation}
It also contains a $\widehat{\mathfrak{u}}(1)_{pk}$ subalgebra generated by the traces $\sum_i K_n^{ii}=\sum_\a L_n^{\a\a}$. For each of these algebras, we can introduce the traceless generators, as in (\ref{slk-constraint}), to define the corresponding $\widehat{\mathfrak{su}}(pk)_1$, $\widehat{\mathfrak{su}}(k)_p$, $\widehat{\mathfrak{su}}(p)_k$ subalgebra. The central charge of the Kac-Moody $\widehat{\mathfrak{su}}(p)_k$ algebra is
\begin{equation}
    c_{p,k}=\frac{k(p^2-1)}{k+p},
\end{equation}
it obeys the identity $c_{pk,1}=c_{k,p}+c_{p,k}$. It implies the following decomposition as a conformal embedding \cite{goddard1985kac, goddard1985symmetric, ALEXANDERBAIS1987561, schellekens1987anomalies},
\begin{equation}\label{conf_red}
    \widehat{\mathfrak{su}}(pk)_1\supset \widehat{\mathfrak{su}}(p)_k\oplus \widehat{\mathfrak{su}}(k)_p.
\end{equation}
In particular, the stress-energy tensors obtained by the Sugawara construction coincide,
\begin{equation}
    T_{\widehat{\mathfrak{su}}(pk)_1}=T_{\widehat{\mathfrak{su}}(p)_k}+T_{\widehat{\mathfrak{su}}(k)_p}\,.
\end{equation}

In \cite{Nakanishi:1990hj}, Nakanishi and Tsuchiya used this decomposition to prove the level-rank duality between $\widehat{\mathfrak{su}}(k)_p$ and $\widehat{\mathfrak{su}}(p)_k$ Kac-Moody algebras. The Fock space $\CF_{\infty}^{(k)}$ describes the states of $pk$ free fermions in the NS sector. It is graded by the fermion number operator $\hat{N_F}=\sum_{r\in\mathbb{Z}}\sum_{i=1}^p\sum_{\alpha=1}^k :\bar\psi^{i,\alpha}_r \psi^{i,\alpha}_{-r}:$ which coincides with the zero mode of the $\widehat{\mathfrak{u}}(1)_{pk}$ current. Accordingly, the Fock space is decomposed into
\begin{equation}
\CF_\infty^{(p,k)}=\bigoplus_{\s\in\mZ}\CF^{(p,k)}_{\infty,\s}.
\end{equation}
We note that the state of our model with $N=mp+q$ belong to the subspace $\CF^{(p,k)}_{\infty,\s}$ with $\s=qk$. For each subspace, we have the decomposition
\begin{equation}\label{dual_decomposition}
    \mathcal{F}^{p,k}_{\sigma}\cong \bigoplus_{\boldsymbol{\lambda}}\mathcal{W}^{\widehat{\mathfrak{su}}(p)_k}_{\boldsymbol{\lambda}}\otimes \mathcal{W}^{\widehat{\mathfrak{su}}(k)_p}_{\boldsymbol{\lambda}'}\otimes \mathcal{W}^{\widehat{\mathfrak{u}}(1)}_{\sigma}\,.
\end{equation}
where the index $\boldsymbol{\lambda}$ labels the irreducible representation of $\widehat{\mathfrak{su}}(p)_k$ This label correspond to the sequence of integers $\lambda_1\geq \lambda_2\geq \cdots \geq \lambda_p$ with the conditions, 
    (i) $\lambda_1-\lambda_p\leq k$,
    (ii) $0\leq \lambda_p<k$,
    (iii) $\sum_{i=1}^p \lambda_i \equiv \sigma$ mod. $pk$.
The sequences $\boldsymbol{\lambda}$ describe the dominant integral weights $(k-\lambda_1+\lambda_p)\Lambda_0+\sum_{j=1}^{p-1}(\lambda_j-\lambda_{j-1})\Lambda_j$, where $\Lambda_j$ is the (affine) fundamental weight of $\widehat{su}(p)_k$. We denoted the corresponding representation space as $\mathcal{W}^{\widehat{\mathfrak{su}}(p)_k}_{\boldsymbol{\lambda}}$. Finally, $\boldsymbol{\lambda}'$ is obtained by transposition of the partition defined by $\boldsymbol{\lambda}$. We refer to section 2 of \cite{Nakanishi:1990hj} for more details, or the original paper \cite{hasegawa1989spin} (see also \cite{altschuler1990branching}, \cite{Walton:1988bs} for the similar results).


The decomposition (\ref{dual_decomposition}) implies an identity for the characters of the corresponding representations. As an illustration, this identity is checked explicitly in appendix \ref{KM22} in the case $p=k=2$.

In the large $N$ limit, we expect that the $\widehat{\mathfrak{su}}(k)_+$-invariance (\ref{slk-constraint}) of the bra state $\bra{\bsx,\bsy}$ is replaced by the condition
\begin{equation}\label{suk-constraint}
    \langle \boldsymbol{x},\boldsymbol{y}|\bar{L}^{\alpha\beta}_n = 0,\quad n\leq 0\,,
\end{equation}
where $\bar{L}^{\alpha\beta}_n$ is the traceless generator of $\widehat{\mathfrak{su}}(k)_p$ obtained from $L^{\alpha\beta}_n$. If so, the decomposition (\ref{conf_red}) implies that the linearly-independent eigenfunctions are spanned by $\widehat{\mathfrak{su}}(p)_k\oplus \widehat{\mathfrak{u}}(1)_{pk}$. The extra $\widehat{\mathfrak{u}}(1)_{pk}$ factor can be introduced by an extra free boson. In this way, we recover the Kac-Moody symmetry of the model observed in \cite{Dorey2016}.

\section{Discussion}
In this paper, we proposed a microscopic description of the non-Abelian FQHE using a spin Calogero Hamiltonian involving $k$th order symmetric representations of the non-Abelian $U(p)$ symmetry. This description follows from the diagonalization of the DTT matrix model \cite{Dorey2016} describing the vortices of a 2+1 dimensional Chern-Simons theory with $U(p)$ gauge group. We have introduced a family of wave functions constructed as determinants involving both coordinates and spin dependence called \textit{wedge states}. We have proved that the Hamiltonian has a triangular action on these wave functions and deduced its spectrum and the expression of ground-state wave functions. The wedge states are not linearly independent when the Chern-Simons level $k$ is greater than one, and it is necessary to quotient by the loop algebra invariance $\widehat{\mathfrak{su}}(k)_+$ to find a proper eigenbasis. The action of this symmetry algebra was obtained by re-expressing the wedge states' wave functions in a free fermion formalism. In this way, it was shown that the $\widehat{\mathfrak{su}}(k)_+$-symmetry produces bilinear relations among determinants, which generalize the celebrated Pl\"ucker relations.

In the large $N$ limit, where $N$ is the number of vortices, the $\widehat{\mathfrak{su}}(k)_+$-invariance is expected to extend to an invariance under the Kac-Moody $\widehat{\mathfrak{su}}(k)_p$ algebra dual to the $\widehat{\mathfrak{su}}(p)_k$-symmetry of the model under level-rank duality. As a result, the spectrum of the Hamiltonian reproduces the WZW $\widehat{\mathfrak{su}}(p)_k$ characters, in agreement with the observations made in \cite{Dorey2016}. In fact, the presence of this Kac-Moody symmetry is already observed at finite $N$. Indeed, it was shown that ground state wave functions obey a Knizhnik-Zamolodchikov equation, generalizing a similar result obtained in \cite{Dorey2016} to the case of degenerate ground states. This proof is independent of our observation of the $\widehat{\mathfrak{su}}(k)_+$-symmetry, and it would be instructive to relate this KZ-equation to the finite $N$ $\widehat{\mathfrak{su}}(k)_+$-primary field condition given in (\ref{KMprimary}).

One of the main open questions is the integrability of our model. The standard spin Calogero model in which particles carry a fundamental representation of the symmetry group $U(p)$ was shown to be integrable using the Lax formalism in \cite{Hikami1993}. In addition, this model exhibits Yangian symmetry \cite{Bernard1994}, and it is natural to expect that our model is also Yangian-invariant, albeit involving higher-order representations. In fact, we expect our Hamiltonian to belong to the class of \textit{degenerate integrable systems} recently introduced by Reshetikhin as a generalization of the idea of quantum integrability (see \cite{Reshetikhin:2015pma, Reshetikhin:2015rba} and the references therein).\footnote{The spin Calogero-Moser model is indeed an example of this class of models.} This notion refers to integrable systems 
in which the invariant phase space has fewer dimensions than the usual ones. Our construction of the wedge states by projection on the bra state described in section 4.1 suggests some similarity with this notion. We hope to be able to come back to this important question in a future publication.

Several studies have indicated a possible role of the $W_{1+\infty}$-algebra and its deformations (e.g., affine Yangian) in the QHE (see, for example, \cite{cappelli1993infinite,Estienne2011}).\footnote{We refer to the review \cite{Matsuo:2023lky} for more details on these algebraic structures.} In this context, several deformations of the model can be introduced, like the trigonometric or elliptic deformations, relativistic deformations, or $\b$-deformations of the coupling. The relativistic model is expected to be diagonalized using the $q$-deformed wedge states introduced in \cite{kashiwara1995decomposition}. Moreover, it would be interesting to investigate the interplay between these algebraic structures and the braiding of non-Abelian anyons.

Finally, from a more physical perspective, it was noted in \cite{Dorey2016} that the model reproduces the Blok-Wen wave functions with filling factor $\nu=p/(k+pn)$ for $n=1$. The KZ equation and Hamiltonian eigevalue equation can accomodate for any $n\geq0$ simply by conjugation with powers of the Vandermonde determinant, but the matrix model interpretation is unclear. It seems important to better understand the physical implication of this manipulation.

\section*{Acknowledgments} JEB would like to thank Dominic Williamson for bringing the article \cite{Dorey2016} to his attention and for many interesting discussions. We wish to thank Didina Serban and Jules Lamers for a critical discussion that led to the understanding of connections with the level-rank duality and with degenerate integrable systems. We also thank Thomas Quella for discussions on the quantum Hall effect, Kantaro Ohmori for explaining the recent realization of non-Abelian anyons on quantum computers, and Y. Fukusumi for the explanation of the recent study of the relation between the Kac-Moody symmetry and the FQHE. Finally, we are thankful to Nick Dorey for carefully reading the manuscript and for many interesting discussions on the relationship between their work and ours. YM is partly supported by KAKENHI Grant-in-Aid No.18K03610, 23K03380 and 21H05190. The authors would like to thank the hospitality of the MATRIX institute where part of this work was carried out.

\appendix

\section{Proof of identities}\label{AppA}
\subsection{Proof of the Hamiltonian action (\ref{trig})}\label{sec:proof}
We rewrite the Hamiltonian as,
\begin{align}\label{H2}
	\CH_\D&=\Delta(\bsx)^{-1}\tilde{\CH}\Delta(\bsx)=N^2+\sum_a(-\partial_a^2+2x_a\partial_a)+\sum_{a\neq b} h_{a,b}\\
 \text{with}\quad h_{a,b}&=-\frac{1}{x_a-x_b}(\partial_a-\partial_b)+\frac{J_{a,b}J_{b,a}}{(x_a-x_b)^2}
\end{align}
We have to prove that $\mathcal{H}_\D$ acting on the product of determinants $\prod_{\alpha=1}^k\mathcal{Y}_{\boldsymbol{r}^{(\a)}}$ produces the diagonal term and the subdominant terms in (\ref{trig}). Since the Hamiltonian is of second order in the derivatives, we can decompose
\begin{align}
    \mathcal{H}_\D \prod_{\alpha=1}^k\mathcal{Y}_{\bsr^{(\a)}} 
    &=
    N^2 \prod_{\alpha=1}^k\mathcal{Y}_{\bsr^{(\a)}}
    \nonumber\\
    &~~~~~+ \sum_\beta\left(\prod_{\alpha(\neq\beta)}\mathcal{Y_{\boldsymbol{r^{(\a)}}}}\right)
    \left(\sum_a(-\partial_a^2+2x_a \partial_a)+\sum_{a\neq b}h_{a,b}\right)\mathcal{Y_{\bsr^{(\b)}}}(x,y)\label{second}\\
    &~~~~~+2\sum_{\beta<\gamma} \left(\prod_{\alpha(\neq \beta,\gamma)}\mathcal{Y}_{\bsr^{(\a)}}\right)
    \left(
    -\sum_a\partial_a \mathcal{Y}_{\bsr^{(\b)}}\partial_a \mathcal{Y}_{\bsr^{(\g)}}
    +\sum_{a\neq b}\frac{1}{(x_a-x_b)^2}J_{a,b}\mathcal{Y}_{\bsr^{(\b)}}J_{b,a}\mathcal{Y}_{\bsr^{(\g)}}
    \right)\label{third}.
\end{align}
The first line is just a multiplication by a constant. The second line gives the action of the derivative on a single determinant $\mathcal{Y}$. The third line gives the second-order derivative acting on distinct determinants.

\paragraph{Computation of the action on a single determinant (\ref{second})}
We omit the index $\beta$ in (\ref{second}) and assume $n_1\leq n_2\leq\cdots \leq n_N$ in $\bsr^{(\b)}=\boldsymbol{r}=\{r_1,\cdots, r_N\}$ with the translation rule $r_a=p n_a+i_a-1$ for $a,b=1,\cdots N$. Using the properties
\begin{align}
\begin{split}
\sum_a x_a\p_a \CY_\bsr(\bsx,\bsy)=|\bsr|\CY_\bsr(\bsx,\bsy),\quad \sum_a \p_a^2\CY_\bsr(\bsx,\bsy)=\sum_a n_a(n_a-1)[x^{n_1}y_{i_1}\wedge\cdots\wedge \underset{\hat{a}}{x^{n_a-2}y_{i_a}}\wedge \cdots \wedge x^{n_N}y_{i_N}]
\end{split}
\end{align}
the terms in (\ref{second}) can be written in the form,
\begin{align}
	\left(\sum_a(-\partial_a^2+2x_a \partial_a)+\sum_{a\neq b}h_{a,b}\right)\mathcal{Y_{\boldsymbol{r}}}(\bsx,\bsy)&= 2|\boldsymbol{r}|\mathcal{Y}_{\bsr}(\bsx,\bsy)\label{first2}\\
	&-\sum_a n_a(n_a-1) [x^{n_1}y_{i_1}\wedge\cdots\wedge \underset{\hat{a}}{x^{n_a-2}y_{i_a}}\wedge \cdots \wedge x^{n_N}y_{i_N}]\label{second2}\\
    & +2\sum_{a<b}h_{a,b} \mathcal{Y}_{\bsr}(\bsx,\bsy)\,.\label{third2}
\end{align}
Eq.(\ref{first2}) defines the diagonal term and the eigenvalue associated with $\mathcal{Y}_{\bsr}$ since $|\bsr|=\sum_a n_a$, and $E(\bsr)=N^2+2\sum_a n_a$. The second line (\ref{second2}) is a sum of wedge states $\mathcal{Y}_{\mathbf{s}}$ with $|\boldsymbol{{s}}|=|\boldsymbol{r}|-2$, which can be included in the subdominant terms in (\ref{trig}). It remains to evaluate the third line (\ref{third2}), and we will show that it produces a sum of wedge states of the form $\mathcal{Y}_{\mathbf{s}}(\bsx,\bsy)$ with $|\mathbf{s}|=|\bsr|-2$. To do so we use an expansion of the determinant with respect to rows $a$ then $b$,
\begin{equation}
\mathcal{Y}_{\boldsymbol{r}}(\bsx,\bsy)=\sum_{c<d}(-)^{a+b+c+d} \mathcal{Y}_{\boldsymbol{r},[ab|cd]}(\bsx,\bsy)\left|
\begin{array}{cc} x_a^{n_c} y_{i_c,a} & x_a^{n_d} y_{i_d,a}\\
    x_b^{n_c} y_{i_c,b} & x_b^{n_d} y_{i_d,b}
    \end{array}
\right|,
\end{equation}
where we took $a<b$ and $\mathcal{Y}_{\boldsymbol{r},[ab|cd]}(\bsx,\bsy)$ denotes the second minor of the $N\times N$ determinant $\CY_\bsr(\bsx,\bsy)$ with respect to rows $a,b$ and columns $c,d$. Since the minors $\mathcal{Y}_{\boldsymbol{r},[ab|cd]}(\bsx,\bsy)$ are independent of the variables $x_a$, $x_b$ and $y_{i,a}$, $y_{i,b}$, the operator $h_{a,b}$ acts only on the two by two determinants, and gives
\begin{align}
    & h_{a,b} \left|\begin{array}{cc} x_a^{n_c} y_{i_c,a} & x_a^{n_d} y_{i_d,a}\\
    x_b^{n_c} y_{i_c,b} & x_b^{n_d} y_{i_d,b}
    \end{array}
    \right|\nonumber\\
    & =- \frac{1}{x_a-x_b}
    \left(
    \left|\begin{array}{cc} n_c x_a^{n_c-1} y_{i_c,a} & n_d x_a^{n_d-1} y_{i_d,a}\\
    x_b^{n_c} y_{i_c,b} & x_b^{n_d} y_{i_d,b}
    \end{array}
    \right|-
    \left|\begin{array}{cc} x_a^{n_c} y_{i_c,a} & x_a^{n_d} y_{i_a,d}\\
    n_c x_b^{n_c-1} y_{i_c,b} & n_d x_b^{n_d-1} y_{i_d,b}
    \end{array}
    \right|
    \right)
    \nonumber \\ 
    & ~~~~+ \frac{1}{(x_a-x_b)^2}\left(
    \left|\begin{array}{cc} x_a^{n_c} y_{i_c,a} & x_a^{n_d} y_{i_d,a}\\
    x_b^{n_c} y_{i_c,b} & x_b^{n_d} y_{i_d,b}
    \end{array}
    \right|
    -
    \left|\begin{array}{cc} x_b^{n_c} y_{i_c,a} & x_b^{n_d} y_{i_d,a}\\
    x_a^{n_c} y_{i_c,b} & x_a^{n_d} y_{i_d,b}
    \end{array}
    \right|
    \right)
\,.\label{hab-action}
\end{align}
where we traded the spin permutation for a coordinate permutation in the second line.

Assuming $n_c<n_d$, we can use the identity,
\begin{align}
	&-\frac{1}{x_a-x_b}(\frac{n_c}{x_a}-\frac{n_d}{x_b}) +\frac{1}{(x_a-x_b)^2}\left(1-(x_a/x_b)^{n_d-n_c}\right)\nonumber\\
	&~~~~~~~ = \frac{n_c}{x_a x_b}-\sum_{l=0}^{n_d-n_c-2}(l+1)x_a^{n_d-n_c-2-l}x_b^{n_c-n_d+l}
\end{align}
to rewrite (\ref{hab-action}) as the sum of determinants,
\begin{align}
    n_c \left|\begin{array}{cc} x_a^{n_c-1} y_{i_c,a} & x_a^{n_d-1} y_{i_d,a}\\
    x_b^{n_c-1} y_{i_c,b} & x_b^{n_d-1} y_{i_d,b}
    \end{array}
    \right|
    -\sum_{l=0}^{n_d-n_c-2} (l+1)
    \left|\begin{array}{cc} x_a^{n_d-2-l} y_{i_c,a} & x_a^{n_c+l} y_{i_d,a}\\
    x_b^{n_d-2-l} y_{i_c,b} & x_b^{n_c+l} y_{i_d,b}
    \end{array}
    \right|.
\end{align}
Re-combining the minors and taking the summation over the indices $a,b$, we arrive at the following sum of wedge states,
\begin{align}
	&n_c [\cdots\wedge x^{n_c-1}y_{i_c}\wedge \cdots\wedge x^{n_d-1}y_{i_d} \wedge\cdots]
	-\sum_{l=0}^{n_d-n_c-2} (l+1) [\cdots\wedge x^{n_d-2-l}y_{i_c}\wedge \cdots\wedge x^{n_c+l}y_{i_d} \wedge \cdots]\,.
\end{align}
All the wedge states in the r.h.s. are indeed of the form $\mathcal{Y}_{\mathbf{s}}(\bsx,\bsy)$ with $|\mathbf{s}|=|\bsr|-2$. The argument in the case $n_c>n_d$ follows the same lines. Finally, the expression vanishes when $n_c=n_d$.

\paragraph{Computation of the cross-terms (\ref{third})} To prove the theorem \ref{trig}, it remains to show that the third line \ref{third} also produces a sum of product of wedge states of total degree $|\mathfrak{r}|-2$. More precisely, introducing the shift operators $T_a=e^{\p_{n_a}}=e^{p\p_{r_a}}$, and the permutation $R_{a,b}^{[\bsr,\bsr']}$ exchanging $r_a\in\bsr$ and $r'_b\in\bsr'$, we will show that
\begin{align}
\begin{split}\label{JYJY}
&\sum_a \p_a\CY_{\bsr}(\bsx,\bsy)\p_a\CY_{\bsr'}(\bsx,\bsy)-\sum_{a\neq b}\dfrac1{(x_a-x_b)^2}(J_{b,a}\CY_{\bsr}(\bsx,\bsy))(J_{a,b}\CY_{\bsr'}(\bsx,\bsy))\\
=&\sum_{c,d}\sum_{\g=0}^{n_c-1}\sum_{\d=0}^{n_d'-1}T_c^{-\g+\d-1}(T_d')^{-\d+\g-1}R_{c,d}^{[\bsr,\bsr']}\CY_{\bsr}(\bsx,\bsy)\CY_{\bsr'}(\bsx,\bsy).
\end{split}
\end{align}
The total degree of the terms entering in the r.h.s. is
\begin{equation}
\left(|\bsr|-\lfloor r_c/p\rfloor+\lfloor r'_d/p\rfloor -\d+\g-1\right)+\left(|\bsr'|-\lfloor r'_d/p\rfloor+\lfloor r_c/p\rfloor -\g+\d-1\right)=|\bsr|+|\bsr'|-2.
\end{equation} 

To prove the identity \ref{JYJY}, we need the lemma,
\begin{equation}\label{lemma1}
J_{b,a}\CY_{\bsr}(\bsx,\bsy)=\sum_{i=1}^p y_{i,b}\p_{i,a}\CY_\bsr(\bsx,\bsy) =\sum_c (-)^{a+c}y_{i_c,b}(x_a^{n_c}-x_b^{n_c})\CY_{\bsr,[a|c]}(\bsx,\bsy).
\end{equation} 
This lemma is obtained by decomposing the determinant by the minors $\CY_{\bsr,[a|c]}(\bsx,\bsy)$ with respect to the row $a$ and column $c$,
\begin{equation}
J_{b,a}\CY_\bsr(\bsx,\bsy)=\sum_i\sum_c(-)^{a+c}y_{i,b}x_a^{n_c}\p_{i,a}(y_{i_c,a})\CY_{\bsr,[a|c]}(\bsx,\bsy)=\sum_c(-)^{a+c}y_{i_c,b}x_a^{n_c}\CY_{\bsr,[a|c]}(\bsx,\bsy).
\end{equation} 
After anti-symmetrization, we have
\begin{equation}
J_{b,a}\CY_\bsr(\bsx,\bsy)=\sum_c(-)^{a+c}y_{i_c,b}(x_a^{n_c}-x_b^{n_c})\CY_{\bsr,[a|c]}(\bsx,\bsy)+\sum_c(-)^{a+c}y_{i_c,b}x_b^{n_c}\CY_{\bsr,[a|c]}(\bsx,\bsy).
\end{equation} 
We then notice that the last term can be re-summed and gives the minor expansion of the determinant $\CY_{\bsr}(\bsx,\bsy)$ in which the column $a$ has been replaced by a copy of the column $b$, and so it is vanishing. The remaining expression gives the r.h.s. of the lemma \ref{lemma1}.

Next, we examine the following quantity, and apply twice the previous lemma to write down
\begin{align}
\begin{split}
&\sum_{a\neq b}\dfrac1{(x_a-x_b)^2}(J_{b,a}\CY_{\bsr}(\bsx,\bsy))(J_{a,b}\CY_{\bsr'}(\bsx,\bsy))\\
=&-\sum_{a\neq b}\sum_{c,d}(-)^{a+b+c+d}y_{i_c,b}y_{i_d',a}\dfrac{x_a^{n_c}-x_b^{n_c}}{x_a-x_b}\dfrac{x_a^{n_d'}-x_b^{n_d'}}{x_a-x_b}\CY_{\bsr,[a|c]}(\bsx,\bsy)\CY_{\bsr',[b|d]}(\bsx,\bsy)\\
=&-\sum_{a\neq b}\sum_{c,d}(-)^{a+b+c+d}y_{i_c,b}y_{i_d',a}\sum_{\g=0}^{n_c-1}\sum_{\d=0}^{n_d'-1}x_a^{n_d'+\g-\d-1}x_b^{n_c-\g+\d-1}\CY_{\bsr,[a|c]}(\bsx,\bsy)\CY_{\bsr',[b|d]}(\bsx,\bsy)\\
=&-\sum_{a,b}\sum_{c,d}(-)^{a+b+c+d}y_{i_c,b}y_{i_d',a}\sum_{\g=0}^{n_c-1}\sum_{\d=0}^{n_d'-1}x_a^{n_d'+\g-\d-1}x_b^{n_c-\g+\d-1}\CY_{\bsr,[a|c]}(\bsx,\bsy)\CY_{\bsr',[b|d]}(\bsx,\bsy)\\
&+\sum_{a}\sum_{c,d}(-)^{c+d}y_{i_c,a}y_{i_d',a}n_cn_d'x_a^{n_c+n_d'-2}\CY_{\bsr,[a|c]}(\bsx,\bsy)\CY_{\bsr',[a|d]}(\bsx,\bsy).
\end{split}
\end{align}
In the last two lines, we completed the sum with the term $a=b$. We recognize in the last term the expansion in minors of the expression
\begin{align}
\begin{split}
\sum_a \p_a\CY_{\bsr}(\bsx,\bsy)\p_a\CY_{\bsr'}(\bsx,\bsy)=\sum_a\left(\sum_{c}(-)^{a+c}y_{i_c,a}n_cx_a^{n_c-1}\CY_{\bsr,[a|c]}(\bsx,\bsy)\right)\left(\sum_{d}(-)^{a+d}y_{i_d',a}n_d'x_a^{n_d'-1}\CY_{\bsr',[a|d]}(\bsx,\bsy)\right)
\end{split}
\end{align}
Thus, by re-arranging the terms, we arrive at 
\begin{align}
\begin{split}
&\sum_a \p_a\CY_{\bsr}(\bsx,\bsy)\p_a\CY_{\bsr'}(\bsx,\bsy)-\sum_{a\neq b}\dfrac1{(x_a-x_b)^2}(J_{b,a}\CY_{\bsr}(\bsx,\bsy))(J_{a,b}\CY_{\bsr'}(\bsx,\bsy))\\
=&\sum_{a,b}\sum_{c,d}\sum_{\g=0}^{n_c-1}\sum_{\d=0}^{n_d'-1}(-)^{a+b+c+d}y_{i_c,b}y_{i_d',a}x_a^{n_d'+\g-\d-1}x_b^{n_c-\g+\d-1}\CY_{\bsr,[a|c]}(\bsx,\bsy)\CY_{\bsr',[b|d]}(\bsx,\bsy).
\end{split}
\end{align}
We can now perform the summations of indices $a$ and $b$; they correspond to the minor expansions of the following determinants 
\begin{align}
\begin{split}
&\sum_a(-)^{a+c}y_{i_d',a}x_a^{n_d'+\g-\d-1}\CY_{\bsr,[a|c]}(\bsx,\bsy)=[x^{n_1}y_{i_1}\wedge\cdots\wedge \underset{\hat{c}}{x^{n'_d+\g-\d-1}y_{i'_d}}\wedge \cdots \wedge x^{n_N}y_{i_N}],\\
&\sum_{b}(-)^{b+d}y_{i_c,b}x_b^{n_c-\g+\d-1}\CY_{\bsr',[b|d]}(\bsx,\bsy)=[x^{n'_1}y_{i'_1}\wedge\cdots\wedge \underset{\hat{d}}{x^{n_c-\g+\d-1}y_{i_c}}\wedge \cdots \wedge x^{n'_N}y_{i'_N}].
\end{split}
\end{align}
The first line corresponds to $\CY_{\bsr}(\bsx,\bsy)$ with $r_c$ replaced by $r'_d+\g-\d-1$, and the second one to $\CY_{\bsr'}(\bsx,\bsy)$ with $r'_d$ replaced by $r_c-\g+\d-1$. These operations have been encoded using the permutation $R_{a,b}^{[\bsr,\bsr']}$ and the shift operator $T_a$ in \ref{JYJY}. It concludes the proof of the Hamiltonian action \ref{trig}.

\subsection{Proof of the Knizhnik–Zamolodchikov equation}\label{A:KP}
In this subsection, we use the following expansion property for rational fractions of the form
\begin{equation}\label{exp_ratio}
\dfrac{x_a^\a x_b^\b-x_a^{\b}x_b^{\a}}{x_a-x_b}=
\begin{cases}
-\sum_{\g=\a}^{\b-1}x_a^{\a+\b-1-\g}x_b^\g & \a<\b,\\
\sum_{\g=\b}^{\a-1}x_a^{\a+\b-1-\g}x_b^\g & \a>\b,
\end{cases},\quad \a,\b\in\mZ^{\geq0}.
\end{equation}

\paragraph{Decomposing the KZ equation} Using the chain rule, we can write
\begin{align}
\begin{split}\label{decomp}
\p_a\Phi_{\rf_0}(\bsx,\bsy)&=\sum_\a\prod_{\b\neq\a}\CY_{\bsr_0^{(\b)}}(\bsx,\bsy)\ \p_a\CY_{\bsr_0^{(\a)}}(\bsx,\bsy),\\
\sum_{b\neq a}\frac{J_{a,b}J_{b,a}}{x_a-x_b}\Phi_{\rf_0}(\bsx,\bsy)&=\sum_\a\prod_{\b\neq\a}\CY_{\bsr_0^{(\b)}}(\bsx,\bsy)\sum_{b\neq a}\frac{J_{a,b}J_{b,a}\CY_{\bsr_0^{(\a)}}(\bsx,\bsy)}{x_a-x_b}\\
+&\sum_{\a\neq\b}\prod_{\g\neq\a,\b}\CY_{\bsr_0^{(\g)}}(\bsx,\bsy)\sum_{b\neq a}\frac{J_{a,b}\CY_{\bsr_0^{(\a)}}(\bsx,\bsy)J_{b,a}\CY_{\bsr_0^{(\b)}}(\bsx,\bsy)}{x_a-x_b}.
\end{split}
\end{align}
In this subsection, we prove the two following identities
\begin{align}
&\sum_{b\neq a}\frac{J_{a,b}J_{b,a}\CY_{\bsr_0}(\bsx,\bsy)}{x_a-x_b}=(p+1)\p_a\CY_{\bsr_0}(\bsx,\bsy),\label{KZ1}\\
&\sum_{b\neq a}\dfrac{J_{b,a}\CY_{\bsr_0}(\bsx,\bsy)J_{a,b}\CY_{\bsr_0'}(\bsx,\bsy)}{x_a-x_b}=\CY_{\bsr_0}(\bsx,\bsy)\p_a\CY_{\bsr_0'}(\bsx,\bsy),\label{KZ2}
\end{align}
from which we deduce
\begin{align}
\begin{split}
&\sum_\a\prod_{\b\neq\a}\CY_{\bsr_0^{(\b)}}(\bsx,\bsy)\sum_{b\neq a}\frac{J_{a,b}J_{b,a}\CY_{\bsr_0^{(\a)}}(\bsx,\bsy)}{x_a-x_b}=(p+1)\sum_\a\prod_{\b\neq\a}\CY_{\bsr_0^{(\b)}}(\bsx,\bsy)\ \p_a\CY_{\bsr_0^{(\a)}}(\bsx,\bsy),\\
&\sum_{\a\neq\b}\prod_{\g\neq\a,\b}\CY_{\bsr_0^{(\g)}}(\bsx,\bsy)\sum_{b\neq a}\frac{J_{a,b}\CY_{\bsr_0^{(\a)}}(\bsx,\bsy)J_{b,a}\CY_{\bsr_0^{(\b)}}(\bsx,\bsy)}{x_a-x_b}=(k-1)\sum_\a\prod_{\b\neq\a}\CY_{\bsr_0^{(\b)}}(\bsx,\bsy)\ \p_a\CY_{\bsr_0^{(\a)}}(\bsx,\bsy).
\end{split}
\end{align}
It shows the KZ relation \ref{eq:KZ} using the decomposition \ref{decomp} for left/right hand sides.

%

\paragraph{Proof of (\ref{KZ1}):} To prove this identity, we first notice that the action of $J_{a,b}J_{b,a}$ on a single determinant $\CY_{\bsr}(\bsx,\bsy)$ can be rewritten using the permutation operator $\t_{a,b}$ exchanging the coordinates $x_a$ and $x_b$,
\begin{equation}
J_{a,b}J_{b,a}\CY_{\bsr}(\bsx,\bsy)=(1-\t_{a,b})\CY_{\bsr}(\bsx,\bsy).
\end{equation} 
To compute its action explicitly, we perform a double minor expansion assuming $a<b$ for definiteness
\begin{equation}
\CY_{\bsr}(\bsx,\bsy)=\sum_{c<d}(-)^{a+b+c+d}\ y_{i_c,a}y_{i_d,b}x_a^{n_c}x_b^{n_d}\CY_{\bsr[a,b|c,d]}(\bsx,\bsy)-\sum_{c>d}(-)^{a+b+c+d}\ y_{i_c,a}y_{i_d,b}x_a^{n_c}x_b^{n_d}\CY_{\bsr[a,b|c,d]}(\bsx,\bsy),
\end{equation} 
where $\CY_{\bsr[a,b|c,d]}(\bsx,\bsy)$ denotes the double minor with rows $a$, $b$ and columns $c$, $d$ removed. From the previous remark, we have
\begin{equation}
\sum_{b\neq a}\frac{J_{a,b}J_{b,a}\CY_{\bsr_0}(\bsx,\bsy)}{x_a-x_b}=A_<-A_>,
\end{equation} 
with
\begin{align}
\begin{split}
&A_<=\sum_{b\neq a}\sum_{c<d}(-)^{a+b+c+d}\ y_{i_c,a}y_{i_d,b}\dfrac{x_a^{n_c}x_b^{n_d}-x_a^{n_d}x_b^{n_c}}{x_a-x_b}\CY_{\bsr[a,b|c,d]}(\bsx,\bsy),\\
&A_>=\sum_{b\neq a}\sum_{c>d}(-)^{a+b+c+d}\ y_{i_c,a}y_{i_d,b}\dfrac{x_a^{n_c}x_b^{n_d}-x_a^{n_d}x_b^{n_c}}{x_a-x_b}\CY_{\bsr[a,b|c,d]}(\bsx,\bsy).
\end{split}
\end{align}
We note that $c<d$ implies $n_c\leq n_d$ if $\CY_{\bsr}(\bsx,\bsy)$ is of the vacuum form \ref{vac_wedge} (i.e. $\bsr\in\mathfrak{R}_0$), and conversely $c>d$ implies $n_c\geq n_d$. Using the expansion formulas \ref{exp_ratio}, we find\footnote{Note that the terms $n_c=n_d$ are indeed vanishing on both sides.}
\begin{align}
\begin{split}
&A_<=-\sum_{b\neq a}\sum_{c<d}\sum_{\g=n_c}^{n_d-1}(-)^{a+b+c+d}\ y_{i_c,a}y_{i_d,b}x_a^{n_c+n_d-1-\g}x_b^\g\CY_{\bsr_0[a,b|c,d]}(\bsx,\bsy),\\
&A_>=\sum_{b\neq a}\sum_{c>d}\sum_{\g=n_d}^{n_c-1}(-)^{a+b+c+d}\ y_{i_c,a}y_{i_d,b}x_a^{n_c+n_d-1-\g}x_b^\g\CY_{\bsr_0[a,b|c,d]}(\bsx,\bsy).
\end{split}
\end{align}

The next step is to perform the summation over $b$ which is interpreted as a minor expansion of the determinant $\CY_{\bsr_0[a|c]}(\bsx,\bsy)$ where some replacement has occured. In the case of $A_<$,\footnote{We need to sum here with the sign $b-1$ since the row $a<b$ has been removed from the determinant $\CY_{\bsr_0[a|d]}(\bsx,\bsy)$.}
\begin{equation}
\sum_{b\neq a}(-)^{b-1+d}y_{i_d,b}x_b^\g\CY_{\bsr_0[a,b|c,d]}(\bsx,\bsy),
\end{equation} 
the elements $y_{i_d,b}x_b^{n_d}$ on column $d$ have been replaced by $y_{i_d,b}x_b^{\g}$ with $\g<n_d$. Since $\CY_{\bsr_0}(\bsx,\bsy)$ is of the form \ref{vac_wedge}, this column is already present in $\CY_{\bsr_0[a|c]}(\bsx,\bsy)$, unless it is the one that has been removed previously, i.e. unless $\g=n_c$ and $i_d=i_c$, in which case we find $\CY_{\bsr[a|d]}(\bsx,\bsy)$. Thus,
\begin{equation}
A_<=\sum_{\superp{c<d}{n_c<n_d}}(-)^{a+d}\ \d_{i_c,i_d}y_{i_d,a}x_a^{n_d-1}\CY_{\bsr_0[a|d]}(\bsx,\bsy).
\end{equation} 
The summation over $c$ can be performed by a counting argument, it produces an extra factor $n_d$,\footnote{Decomposing $N=mp+q$, $d=\bar d p+d'$, $c=\bar c p+c'$ with $c',d'\in[\![0,p-1]\!]$. In the vacuum case $\bsr\in\mathfrak{R}_0$, we have $n_c=\bar c$, $n_d=\bar d$, and $i_c=c'+1$, $i_d=d'+1$ for $n_c,n_d<m$ (but arbitrary for $n_c=m$ or $n_d=m$). The restrictions on the summation impose $\bar c=0\cdots \bar d-1$ and $c'=d'$.}
\begin{equation}\label{Alow}
A_<=\sum_{d}(-)^{a+d}\ y_{i_d,a}n_dx_a^{n_d-1}\CY_{\bsr_0[a|d]}(\bsx,\bsy)=\p_a\CY_{\bsr_0}(\bsx,\bsy).
\end{equation} 

Next, we examine $A_>$, and interpret again the summation over $b$ as a minor expansion of the determinant $\CY_{\bsr_0[a|c]}(\bsx,\bsy)$ where the elements $y_{i_d,b}x_b^{n_d}$ on column $d$ have been replaced by $y_{i_d,b}x_b^{\g}$. The only difference is that now $\g\in[\![n_d,n_c-1]\!]$ with $n_c\geq n_d$. In general, the determinant vanishes since the same column is present twice, unless $\g=n_d$ in which case we simply recover the original determinant,
\begin{equation}
\sum_{b\neq a}(-)^{b-1+d}y_{i_d,b}x_b^{n_d}\CY_{\bsr_0[a,b|c,d]}(\bsx,\bsy)=\CY_{\bsr_0[a|c]}(\bsx,\bsy).
\end{equation} 
Thus, we have
\begin{equation}
A_>=-\sum_{\superp{c>d}{n_c>n_d}}(-)^{a+c}\ y_{i_c,a}x_a^{n_c-1}\CY_{\bsr_0[a|c]}(\bsx,\bsy).
\end{equation} 
The sum over $d$ is performed by a similar counting argument which gives a factor $pn_c$ since the condition $\d_{i_c,i_d}$ is not present in this case.\footnote{Using the same decomposition as in the previous argument, we have $\bar d=0\cdots \bar c-1$ and $d'=0\cdots p-1$ since $\bar d<m$.} We find
\begin{equation}\label{Aup}
A_>=-\sum_{c}(-)^{a+c}\ y_{i_c,a}p n_cx_a^{n_c-1}\CY_{\bsr_0[a|c]}(\bsx,\bsy)=-p\p_a\CY_{\bsr_0}(\bsx,\bsy).
\end{equation} 
Combining the results \ref{Alow} and \ref{Aup} for $A_>$ and $A_<$ respectively, we deduce that $A_<-A_>$ produces indeed $(p+1)\p_a\CY_{\bsr_0}(\bsx,\bsy)$, which shows the identity \ref{KZ1}.

\paragraph{Proof of the equation \ref{KZ2}} In this proof, we need the following lemma which can be derived using a similar argument as the previous lemma \ref{lemma1},
\begin{equation}\label{lemma2}
J_{b,a}\CY_{\bsr}(\bsx,\bsy)J_{a,b}\CY_{\bsr'}(\bsx,\bsy)=\sum_{c,d}(-1)^{a+b+c+d}y_{i_c,b}y_{i'_d,a}(x_a^{n_c}x_b^{n'_d}-x_a^{n'_d}x_b^{n_c})\CY_{\bsr[a|c]}(\bsx,\bsy)\CY_{\bsr'[b|d]}(\bsx,\bsy),
\end{equation} 
for $a\neq b$. The proof of this lemma is obtained using a minor expansion of both determinants to compute explicitly the action of $J_{a,b}$,
\begin{equation}
J_{b,a}\CY_{\bsr}(\bsx,\bsy)J_{a,b}\CY_{\bsr'}(\bsx,\bsy)=\sum_{c,d}(-1)^{a+b+c+d}y_{i_c,b}y_{i'_d,a}x_a^{n_c}x_b^{n'_d}\CY_{\bsr[a|c]}(\bsx,\bsy)\CY_{\bsr'[b|d]}(\bsx,\bsy).
\end{equation} 
This expression is then anti-symmetrized to produce \ref{lemma2}, noticing that
\begin{equation}
\sum_{c,d}(-1)^{a+b+c+d}y_{i_c,b}y_{i'_d,a}x_a^{n'_d}x_b^{n_c}\CY_{\bsr[a|c]}(\bsx,\bsy)\CY_{\bsr'[b|d]}(\bsx,\bsy)=0
\end{equation} 
since it corresponds to the minor expansion of determinants with two equal rows (two copies of the row $b$ in the first one, and two copies of the row $a$ in the second one).

Thus, using the lemma \ref{lemma2}, we can write
\begin{equation}
\dfrac{J_{b,a}\CY_{\bsr}(\bsx,\bsy)J_{a,b}\CY_{\bsr'}(\bsx,\bsy)}{x_a-x_b}=\sum_{c,d}(-1)^{a+b+c+d}y_{i_c,b}y_{i'_d,a}\dfrac{x_a^{n_c}x_b^{n'_d}-x_a^{n'_d}x_b^{n_c}}{x_a-x_b}\CY_{\bsr[a|c]}(\bsx,\bsy)\CY_{\bsr'[b|d]}(\bsx,\bsy).
\end{equation} 
Next, we need to perform the summation over $b\neq a$. Assuming $n_c<n'_d$, we can expand the rational function as in \ref{exp_ratio} and obtain
\begin{align}
\begin{split}
\sum_{b\neq a}(-)^{b+d}y_{i_c,b}\dfrac{x_a^{n_c}x_b^{n'_d}-x_a^{n'_d}x_b^{n_c}}{x_a-x_b}\CY_{\bsr'[b|d]}(\bsx,\bsy)=&-\sum_{\g=n_c}^{n'_d-1}x_a^{n_c+n'_d-1-\g}\sum_b(-)^{b+d}y_{i_c,b}x_b^\g\CY_{\bsr'[b|d]}(\bsx,\bsy)\\
&+(n'_d-n_c)(-)^{a+d}y_{i_c,a}x_a^{n_c+n'_d-1}\CY_{\bsr'[a|d]}(\bsx,\bsy),
\end{split}
\end{align}
where the second line corresponds to the extra term $b=a$ added to the summation. The first line contains minor expansions of the determinant $\CY_{\bsr'}(\bsx,\bsy)$ in which the elements $y_{i'_d,b}x_b^{n'_d}$ of the row $d$ have been replaced by $y_{i_c,b}x_b^\g$ with $\g<n'_d$. If $\CY_{\bsr'}(\bsx,\bsy)$ is a vacuum wedge state of the form \ref{vac_wedge}, i.e. $\bsr'\in\mathfrak{R}_0$, this determinant is vanishing and so
\begin{equation}\label{ncnd}
\sum_{b\neq a}(-)^{b+d}y_{i_c,b}\dfrac{x_a^{n_c}x_b^{n'_d}-x_a^{n'_d}x_b^{n_c}}{x_a-x_b}\CY_{\bsr'_0[b|d]}(\bsx,\bsy)=(n'_d-n_c)(-)^{a+d}y_{i_c,a}x_a^{n_c+n'_d-1}\CY_{\bsr_0'[a|d]}(\bsx,\bsy),
\end{equation} 
This equation is also trivially satisfied if $n_c=n'_d$.

When $n_c>n'_d$, we need to use another expansion for the ratio \ref{exp_ratio}, and we find
\begin{align}
\begin{split}
\sum_{b\neq a}(-)^{b+d}y_{i_c,b}\dfrac{x_a^{n_c}x_b^{n'_d}-x_a^{n'_d}x_b^{n_c}}{x_a-x_b}\CY_{\bsr'[b|d]}(\bsx,\bsy)=&\sum_{\g=n'_d}^{n_c-1}x_a^{n_c+n'_d-1-\g}\sum_b(-)^{b+d}y_{i_c,b}x_b^\g\CY_{\bsr'[b|d]}(\bsx,\bsy)\\
&+(n'_d-n_c)(-)^{a+d}y_{i_c,a}x_a^{n_c+n'_d-1}\CY_{\bsr'[a|d]}(\bsx,\bsy),
\end{split}
\end{align}
The first line still contains minor expansions of the determinant $\CY_{\bsr'}(\bsx,\bsy)$ in which the elements $y_{i'_d,b}x_b^{n'_d}$ of the row $d$ have been replaced by $y_{i_c,b}x_b^\g$, but now $n'_d\leq\g\leq n_c-1$. If both $\bsr,\bsr'\in\mathfrak{R}_0$, then $0\leq n_c, n'_d\leq m$ and so $\g<m$. Thus, this row will be already present in $\CY_{\bsr_0'[b|d]}(\bsx,\bsy)$ unless it is the one that had been removed from $\CY_{\bsr_0'}(\bsx,\bsy)$, i.e. unless $\g=n'_d$ and $i_c=i'_d$. In this case, we recover the original determinant,
\begin{equation}\label{ndnc}
\sum_{b\neq a}(-)^{b+d}y_{i_c,b}\dfrac{x_a^{n_c}x_b^{n'_d}-x_a^{n'_d}x_b^{n_c}}{x_a-x_b}\CY_{\bsr_0'[b|d]}(\bsx,\bsy)=\d_{i_c,i'_d}x_a^{n_c-1}\CY_{\bsr'_0}(\bsx,\bsy)+(n'_d-n_c)(-)^{a+d}y_{i_c,a}x_a^{n_c+n'_d-1}\CY_{\bsr_0'[a|d]}(\bsx,\bsy),
\end{equation}

Combining the results \ref{ncnd} and \ref{ndnc}, we find
\begin{align}
\begin{split}
\sum_{b\neq a}\dfrac{J_{b,a}\CY_{\bsr_0}(\bsx,\bsy)J_{a,b}\CY_{\bsr_0'}(\bsx,\bsy)}{x_a-x_b}=&\sum_{\superp{c,d}{n_c>n'_d}}(-)^{a+c}\d_{i_c,i'_d}x_a^{n_c-1}y_{i'_d,a}\CY_{\bsr_0[a|c]}(\bsx,\bsy)\CY_{\bsr'_0}(\bsx,\bsy)\\
&+\sum_{c,d}(-1)^{c+d}(n'_d-n_c)y_{i_c,a}y_{i'_d,a}x_a^{n_c+n'_d-1}\CY_{\bsr_0[a|c]}(\bsx,\bsy)\CY_{\bsr_0'[a|d]}(\bsx,\bsy).
\end{split}
\end{align}
The summation over $d$ in the first line can be simplified by a counting argument. Examining the values of $i'_d$ and $n'_d$ for $a=1\cdots N$, we find a factor $n_c$,\footnote{Note that the last values $n'_d=m$ and $i'_d=s_\a$ are never reached thanks to the condition $n'_d<n_c\leq m$.}
\begin{center}
\begin{tabular}{c|ccccccccccccccccc}
$d$ & $1$ & $\cdots$ & $p$ & $p+1$ & $\cdots$ & $2p$ & $2p+1$ & $\cdots$ & $3p$ & $\cdots$ & $\cdots$ & $(m-1)p+1$ & $\cdots$ & $mp$ & $mp+1$ & $\cdots$ & $N$\\
\hline
$i'_d$ & $1$ & $\cdots$ & $p$ & $1$ & $\cdots$ & $p$ & $1$ & $\cdots$ & $p$ & $\cdots$ & $\cdots$ & $1$ & $\cdots$ & $p$ & $s_1$ & $\cdots$ & $s_q$\\
$n'_d$ & $0$ & $\cdots$ & $0$ & $1$ & $\cdots$ & $1$ & $2$ & $\cdots$ & $2$ & $\cdots$ & $\cdots$ & $m-1$ & $\cdots$ & $m-1$ & $m$ & $\cdots$ & $m$\\
\end{tabular}
\end{center}
Thus, we find after cancellation and reforming the determinant
\begin{align}
\begin{split}
\sum_{b\neq a}\dfrac{J_{b,a}\CY_{\bsr_0}(\bsx,\bsy)J_{a,b}\CY_{\bsr_0'}(\bsx,\bsy)}{x_a-x_b}=&\sum_{c}(-)^{a+c}n_cx_a^{n_c-1}y_{i_c,a}\CY_{\bsr_0[a|c]}(\bsx,\bsy)\CY_{\bsr'_0}(\bsx,\bsy)\\
&+\sum_{c,d}(-1)^{c+d}(n'_d-n_c)y_{i_c,a}y_{i'_d,a}x_a^{n_c+n'_d-1}\CY_{\bsr_0[a|c]}(\bsx,\bsy)\CY_{\bsr_0'[a|d]}(\bsx,\bsy)\\
=&\sum_{c,d}(-1)^{c+d}n'_dy_{i_c,a}y_{i'_d,a}x_a^{n_c+n'_d-1}\CY_{\bsr_0[a|c]}(\bsx,\bsy)\CY_{\bsr_0'[a|d]}(\bsx,\bsy)\\
=&\CY_{\bsr_0}(\bsx,\bsy)\p_a\CY_{\bsr_0'}(\bsx,\bsy).
\end{split}
\end{align}
This is indeed the identity \ref{KZ2}, which concludes the proof of the KZ equation.

\section{Explicit form of solutions for \texorpdfstring{$N=2$}{N=2}}\label{a:N=2}
In this appendix, we examine the eigenvalue problem for the Hamiltonian \ref{Htilde_m} in the case of two particles ($N=2$) for generic $p$ and $k$.
\paragraph{Spin wave function} The spin states of each particle $a=1,2$ take the form $y_{i_1,a}\cdots y_{i_k,a}$. They belong to the order $k$ totally symmetric representation of $U(p)$ associated to the Young diagram $[k]$. The tensor product of two representations $[k]$ has the decomposition,
\begin{equation}\label{irrepdcmp}
[k]\otimes [k]=[2k,0]\oplus [2k-2,2]\oplus \cdots \oplus [k,k]=\oplus_{l=0}^k [k+l,k-l]\,.
\end{equation}

The spin component of the wave function expressed in terms of $y$-variables is obtained from a function $Y^{HW}_{\boldsymbol{m}}(\bsy)$ satisfying the highest weight condition
\begin{equation}\label{HWC}
K_{i,j}Y^{HW}_{\boldsymbol{m}}(\bsy)=0,\quad i<j,\quad
K_{i,i}Y^{HW}_{\boldsymbol{m}}(\bsy)=m_iY^{HW}_{\boldsymbol{m}}(\bsy),\quad i=1,\cdots, p\,,
\end{equation}
for the $U(p)$ generators $K_{i,j}$ defined in \ref{def_Kij}. In particular, the highest weight state for the irreducible representation $[k+l,k-l]$ has the wave function
\begin{equation}\label{HWSN=2}
	Y^{HW}_{[k+l,k-l]}(\bsy)=(y_{11}y_{22}-y_{12}y_{21})^{k-l}(y_{11})^l(y_{12})^l.
\end{equation}
This state has parity $(-1)^{k-l}$ under the exchange of the two particles $1\leftrightarrow 2$. It satisfies the conditions (\ref{HWC}) with $m_1=k+l,m_2=k-l, m_3=\cdots=m_p=0$.

The operator $J_{12}J_{21}$ can be written using the second Casimir operator of $U(p)$. It is constant on the states that belong to a given representation $[k+l,k-l]$. Its eigenvalue can be obtained from the action on the highest weight state of this representation,
\begin{equation}
	(J_{12}J_{21})Y^{HW}_{[k+l,k-l]}(\bsy)=l(l+1)Y^{HW}_{[k+l,k-l]}(\bsy)\,.
\end{equation}
We will show below that the representation with $l=0$ (i.e., $[k,k]$) has minimal energy and corresponds to the true ground state. Instead, states in the representation $[k+l,k-l]$ with $l=1,\cdots, k$ enter in the wave function of excited states.

We note that there are different choices for the spin components of the wave functions at fixed $l$. In some references on the spin Calogero model (see, e.g., \cite{minahan1993integrable}), a totally symmetric wave function $l=k$ is used. The form of the KZ equation for the vacuum wave functions depends explicitly on this choice of spin component (see, for example, \cite{Hikami1993, Hikami1993a}).

Finally, we comment on the dimension of the representation. The dimension of the symmetric representation of order $k$ is,
\begin{equation}
    d_{p,k}=\frac{\Gamma(k+p)}{\Gamma(k+1)\Gamma(p)}\,.
\end{equation}
On the other hand, the dimension of the irreducible representation $[k+l,k-l]$ is given by 
the factors over hooks rule (see for example, p120 in \cite{georgi2000lie}),
\begin{equation}\label{mpkl}
	\mathfrak{m}(p,k,l) =\frac{(k+p+l-1)!(k+p-l-2)!(2l+1)}{(k+l+1)!(k-l)! (p-1)!(p-2)!}.
\end{equation}
We note that $d_{p,k}^2=\sum_{l=0}^k \mathfrak{m}(p,k,l)$ in agreement with the branching rule (\ref{irrepdcmp}).

\paragraph{Inclusion of the coordinate dependence} The total wave functions can be factorized into coordinate and spin dependence, $\psi(\bsx,\bsy)=X_{l}(\bsx)Y_{l}(\bsy)$
where $Y_{l}(\bsy)$ is assumed to belong to the representation $[k+l,k-l]$. In this case, the Hamiltonian \ref{Htilde_m} acting on the coordinates component $X_{l}(\bsx)$ reduces to
\begin{align}
\tilde{\CH}_{|\bsx}  &= \sum_{a=1,2}(-\partial_a^2+x_a\partial_a)+2+\frac{2l(l+1)}{(x_1-x_2)^2}\,.\label{Htilde}
\end{align}
We recover here the Calogero Hamiltonian, the ground state is found using the ansatz $X_l(\bsx)=(x_1-x_2)^n$. The eigenvalue problem has a solution only if $n=l+1$ or $n=-l$. Only solutions with $n$ positive, i.e., $n=l+1$, are nonsingular. The corresponding eigenvalue is $E_l=2l+4$, and the representation $[k+l,k-l]$ with $l=0$ has indeed the minimal energy.

Combining spin and coordinate components, we deduce the ground state wave functions for $l$ fixed,
\begin{align}
\psi(\bsx,\bsy) & = (x_1-x_2)^{l+1} (y_{11}y_{22}-y_{12}y_{21})^{k-l}y_{11}^ly_{12}^l\\
&=(x_1-x_2)\left|
\begin{array}{cc}
y_{11} & y_{12}\\ y_{21} & y_{22}
\end{array}
\right|^{k-l}
\left|
\begin{array}{cc}
y_{11} & y_{12}\\ y_{11}x_1 & y_{12} x_2
\end{array}
\right|^l
\end{align}
This expression agrees with the formula \ref{psil} for $N=2$. We note that for $k$ even (resp. odd), the system is fermionic (resp. bosonic) under the exchange of the two particles, i.e. the simultaneous exchange of spin and coordinate variables $(x_1,y_{i,1})\leftrightarrow(x_2,y_{i,2})$.


For the singular branch of solutions obtained by choosing $n=-l$, the corresponding energy is  $E_l=-2l+2$. The lowest energy among the representations $[k+l,k-l]$ is found for $l=k$ which gives $E_k=-2k+2$, and the wave function has a pole of order $(x_1-x_2)^{-k}$ which grows with $k$.

	
\paragraph{Excited states} The wave functions of excited states have a coordinate dependence of the form $X_l(\bsx)=x_{12}^{l+1} P(\bsx)$ where $P(\bsx)$ is a symmetric polynomial in the variables $(x_1,x_2)$. After conjugation, the action of the Hamiltonian \ref{Htilde_m} on $P(\bsx)$ takes the form
\begin{equation}
	\tilde{\CH}_{|P}=x_{12}^{-l-1}\tilde{\CH}_{|\bsx} x_{12}^{l+1} = 2l +4 +\sum_a(-\partial_a^2+x_a \partial_a) -\frac{2(l+1)}{x_{12}} (\partial_1-\partial_2).
\end{equation}
The polynomial $P(\bsx)$ can be decomposed on the basis of symmetric monomials $m_{r,s}(\bsx) = x_1^rx_2^s+x_1^s x_2^r$ with $r\geq s\geq 0$.
The action of the Hamiltonian $\tilde{\CH}_{|P}$ on the monomials reads
\begin{equation}
	\tilde{\CH}_{|P}m_{r,s}(\bsx)=2(r+s+l+2) m_{r,s}(\bsx)+\cdots
\end{equation}
where $\cdots$ is a linear combination of $m_{r',s'}(\bsx)$ with $r'+s'=r+s-2$. We recover here the triangular action of the Hamiltonian written in \ref{trig} for general $N$. For each value $E_{r,s}=2(l+r+s+2)$ of the energy, it is possible to construct the corresponding eigenfunction as a linear combination
\begin{equation}
\psi_{l,r,s}(\bsx,\bsy)=Y_l(\bsy)\left(m_{r,s}(\bsx)+\sum_{r'+s'\leq r+s-2} c_{r's'}m_{r',s'}(\bsx)\right).    
\end{equation}
From the formula (\ref{mpkl}) for the dimension of irreducible representations, we deduce the partition function for a system of two particles,
\begin{equation}
    \mbox{Tr}(q^{\tilde{H}}) =q^4 \sum_{l=0}^k \mathfrak{m}(p,k,l) \frac{q^{2l}}{(1-q^2)(1-q^4)}\,.
\end{equation}



\section{Classification of eigenstates using Gelfand-Zetlin patterns}\label{app:GZ}
In this appendix, we focus on the case $k=1$ and describe the decomposition of the wedge state according to irreducible representations of the Lie algebra $u(p)$, or $\gl_p$ upon complexification. This technique is well-known in the context of the spin Calogero model \cite{Uglov1997}. We apply here similar ideas to the wedge states defined in (\ref{Yl}). We note that the Gelfand-Zetlin framework also works for $k>1$, but it requires the decomposition of a tensor product of $k$ irreducible representations, which is difficult in practice.

Irreducible highest weight finite-dimensional representations of $\gl_p$ are labeled by sequences of integers \cite{raczka1986theory},
\begin{equation}
    \lambda_1\geq \lambda_2\geq\cdots \geq\lambda_p,.
\end{equation}
The basis of these irreducible representations is labeled by Gel'fand-Zetlin (GZ) patterns, i.e., a set of integers filling the following table,
\begin{equation}
	\Lambda = \left(
	\begin{array}{ccccccccc}
		\lambda_{1,1} & & \lambda_{2,1} & & \cdots & & \lambda_{p-1,1} & & \lambda_{p,1}\\
		& \lambda_{1,2} & & \lambda_{2,2} &\cdots & & & \lambda_{p-1,2} & \\
		& & \ddots & & \vdots & & \udots & &\\
		& & & \lambda_{1,p-1} & & \lambda_{2,p-1} & & &\\
		& & & & \lambda_{1,p} & & & &
	\end{array}
	\right)
\end{equation}
with the identification $\lambda_{i,1}=\lambda_i$ and the requirement that other integers satisfy $\lambda_{i,j}\geq \lambda_{i,j+1} \geq \lambda_{i+1,j}$. These patterns are in one-to-one correspondence with the sequence of embedding $\mathfrak{gl}_p\supset \mathfrak{gl}_{p-1}\supset\cdots \supset \mathfrak{gl}_1$. Indeed, the integers $\lambda_{i,j}$ ($i=1,\cdots,p-j+1$) correspond to the highest weights of the representation of the subalgebra $\mathfrak{gl}_{p-i+1}$ embedded in the representation of $\mathfrak{gl}_{p-i+2}$ with highest weights $\lambda_{i,j+1}$. The highest weight state of $\gl_p$ corresponds to the Gelfand -Zetlin label $\Lambda_0$ defined by $\lambda_{i,j}=\lambda_{i}$ for any $i,j$.

By construction, the wedge state belongs to a tensor product of $N$ fundamental representations of $\gl_p$. This tensor product decomposes as a sum of irreducible representations labeled by integers $\l_i$ with the constraints
\begin{equation}
    \sum_{i=1}^p \lambda_i =N\,,\quad
    \lambda_p\geq 0\,.
\end{equation}
Thus, the labels $\l_i$ define a partition $\lambda=[\lambda_1,\cdots, \lambda_l]$ of $N$ with $l=\ell(\l)\leq p$. 

Let $\boldsymbol{n}=(n_1,\cdots, n_{h})$ be a set of mutually distinct non-negative integers. The highest weight states of the irreducible representations of $\gl_p$ mentioned before correspond to wedge states written in two equivalent forms,
\begin{align}
    \CY_{[\Lambda_0,\boldsymbol{n}]}(\bsx,\bsy)&:=[y_1x^{n_1}\wedge y_1 x^{n_2}\wedge\cdots \wedge y_1x^{n_{\lambda_1}}\wedge y_2 x^{n_1} \wedge \cdots \wedge y_2 x^{n_{\lambda_2}}\wedge \cdots\wedge y_{l} x^{n_1}\wedge\cdots \wedge y_{l} x^{n_{\lambda_l}}]\label{wedge1}\\
        &=\e [y_1x^{n_1}\wedge y_2 x^{n_1}\cdots \wedge y_{\lambda_1'}x^{n_1}\wedge y_1x^{n_2}\wedge\cdots\wedge y_{\lambda'_2}x^{n_2}\wedge \cdots\wedge y_1 x^{n_{\lambda_1}}\wedge y_{\lambda'_{\lambda_1}} x^{n_{\lambda_1}} ],
        \label{wedge2}
\end{align} 
where $\e$ is a sign of re-ordering the factors in the wedge product, and $\l'=[\l'_1,\cdots,\l'_{l_1}]$ is the transposed partition.
The first form (\ref{wedge1}) is convenient for discussing the highest weight representation given below. The second form (\ref{wedge2}) is useful to derive the energy of the wedge state. 
These two expressions illustrate the $U(N)/U(p)$ duality of the spin Calogero system. 

These states satisfy the highest weight condition with weight $\lambda$,
\begin{equation}
    K_{i,j}\CY_{[\Lambda_0,\boldsymbol{n}]}(\bsx,\bsy)=0,\quad i<j,\quad
    K_{i,i}\CY_{[\Lambda_0,\boldsymbol{n}]}(\bsx,\bsy)=\lambda_i \CY_{[\Lambda_0,\boldsymbol{n}]}(\bsx,\bsy),
    \label{cYlambda}
\end{equation}
where the action of the $\gl_p$ generators $K_{i,j}$ has been defined in \ref{def_Kij}. The highest weight state with the lowest energy in each representation $\l$ is obtained by taking $\boldsymbol{n}=\boldsymbol{n}_0=\left\{0,1,\cdots,\lambda_l-1\right\}$. The corresponding wedge state is an eigenstate of the Hamiltonian \ref{Htilde_m} with energy $E(\lambda)=N^2+\sum_{i=1}^l \lambda_i(\lambda_i-1)$. A short proof of these statements is given below.

\begin{proof}
We note that the wedge state can be written in the form
\begin{equation}
\CY_{[\Lambda_0,\boldsymbol{n}]}(\bsx,\bsy)=\sum_{\s\in S_n}(-)^\s\prod_{i=1}^l\prod_{\a=1}^{\l_i}y_{i,\s(a_{i,\a})}x_{\s(a_{i,\a})}^{n_\a},
\end{equation}
with $a_{i,\a}=\a+\sum_{j=1}^{i-1}\l_i$ labeling the $N$ boxes of the Young diagram $\l$.

\noindent (i) The first condition comes from the fact that $K_{i,j} (y_{k,a} x_a^n) = \delta_{j,k}\ y_{i,a} x_a^n$, but when $i<j$ the same factor exists in the wedge product due to the condition $\lambda_i\leq \lambda_j$.
The expression for the action of $K_{i,i}$ comes from the fact that there are $\lambda_i$ particles of spin $i$, i.e. there are $\l_i$ factors $y_{i,a}$ in each term.

\noindent (ii) The lowest energy state is obtained by minimizing $|\bsr|$ in each representation $\l$, since the Hamiltonian has a global $\gl_p$-invariance. For $\CY_{[\Lambda_0,\boldsymbol{n}]}(\bsx,\bsy)$, we have from (\ref{wedge2}),
\begin{equation}
|\bsr|=\sum_{\a=1}^{\l_1}n_\a \l'_\a\,.
\end{equation}
It is indeed minimized by $n_\a=\a-1$ which gives  
\begin{equation}
|\bsr|=\sum_{\a=1}^{\l_1}(\a-1)\l'_\a=\sum_{i=1}^l\sum_{j=1}^{\l_i}(j-1)=\hf\sum_{i=1}^l \l_i(\l_i-1).
\end{equation}


\end{proof}

\paragraph{Examples}
\begin{itemize}
    \item For the totally symmetric representation $\lambda_1=N, \lambda_2=\cdots=\lambda_p=0$, $l=1$ and $\boldsymbol{n}_0=\{0,1,\cdots, N-1\}$. The corresponding wedge state is proportional to the Vandermonde determinant:
    \begin{equation}
        \CY_{[\Lambda_0,\boldsymbol{n}_0]}(\bsx,\bsy) =[y_1\wedge y_1 x\wedge \cdots \wedge y_1 x^{N-1}] =\left(\prod_{a=1}^N y_{1,a}\right) \Delta(\bsx)\,.\label{sym_rep}
    \end{equation}
    Generic states of the  representation $\l$ of $\gl_p$ are obtained by the application of $K_{i,j}$ which does not modify the $x$-dependence. The Gelfand-Zetlin patterns of this representation have the form,
    \begin{equation}
	\Lambda = \left(
	\begin{array}{ccccccccc}
		N & & 0 & & \cdots & & 0 & & 0\\
		& \lambda_{1,2} & & 0 &\cdots & & &0 & \\
		& & \ddots & & \vdots & & \udots & &\\
		& & & \lambda_{1,p-1} & & 0 & & &\\
		& & & & \lambda_{1,p} & & & &
	\end{array}
	\right),\quad
 0\leq \lambda_{1,p}\leq \lambda_{1,p-1}\leq \cdots \leq \lambda_{1,2}\leq N\,.
\end{equation}
The number of patterns is $\frac{(N+p-1)!}{(p-1)!N!}$, which coincides with the dimension of the representation $[N]$ of $\gl_p$.

    \item In the totally anti-symmetric case with $N\leq p$, we have $\lambda=[\overbrace{1,\cdots,1}^N]$ and $\boldsymbol{n}_0=\{0\}$. The corresponding wedge state does not depend on the variables $x_a$,
    \begin{equation}
        \CY_{[\Lambda_0,\boldsymbol{n}_0]}(\bsx,\bsy)=[y_1\wedge y_2\wedge\cdots \wedge y_N]\,.
    \end{equation}
    The number of patterns for the representation $\l=[\overbrace{1,\cdots,1}^N,\overbrace{0,\cdots,0}^{N-p}]$ is $\frac{p!}{(p-N)!N!}$, which equals to the dimension of the order $N$ totally-antisymmetric representation of $\gl_p$.
    \item With the exception of the two previous cases, in general the coordinate and spin dependence mix. For instance, taking $\lambda=[\lambda_1,\lambda_2]$ with $\l_1+\l_2=N$, we have $\boldsymbol{n}=\{0,1,\cdots, \lambda_1-1\}$ and 
    \begin{equation}
        \CY_{[\Lambda_0,\boldsymbol{n}_0]}(\bsx,\bsy) =\sum_{A^{(1)},A^{(2)}}\epsilon^{A^{(1)} A^{(2)}} \prod_{i=1}^{\lambda_1} y_{1,a_{1i}}\prod_{j=1}^{\lambda_2} y_{2,a_{2j}}
        \Delta(\bsx^{(1)})\Delta(\bsx^{(2)})\,,
    \end{equation}
    where $A^{(1)}=\left\{a_{1,1},\cdots,a_{1,\lambda_1}\right\}$ and $A^{(2)}=\left\{a_{2,1},\cdots,a_{2,\lambda_2}\right\}$ are disjoint division of set of integers $\left\{1,\cdots, N\right\}$, namely $A^{(1)},A^{(2)}\subset \left\{1,\cdots, N\right\}$, $A^{(1)}\cup A^{(2)}=\left\{1,\cdots, N\right\}$, $A^{(1)}\cap A^{(2)}=\emptyset$. 
    $\bsx^{(1)}=\left\{x_{a_{1,1}},\cdots, x_{a_{1,\lambda_1}}\right\}$ and $\bsx^{(2)}=\left\{x_{a_{2,1}},\cdots, x_{a_{2,\lambda_2}}\right\}$\,. The patterns have the form
    \begin{equation}
	\Lambda = \left(
	\begin{array}{ccccccccc}
		\lambda_{1} & & \lambda_{2} & & \cdots & & 0 & & 0\\
		& \lambda_{1,2} & & \lambda_{2,2} &\cdots & & & 0 & \\
		& & \ddots & & \vdots & & \udots & &\\
		& & & \lambda_{1,p-1} & & \lambda_{2,p-1} & & &\\
		& & & & \lambda_{1,p} & & & &
	\end{array}
	\right)
\end{equation}
where all elements but $\lambda_{i,1}$, $\lambda_{i,2}$ ($i=2,\cdots,p$) vanish.
\end{itemize}

\section{Character decomposition and level-rank duality}\label{KM22}
In this appendix, we check the decomposition (\ref{dual_decomposition}) of the fermionic Fock space at the level of characters for $p=k=2$. In this case, irreducible representations are labeled by the integer $l=0,1,2$. The relevant decompositions for the DTT matrix model are
\begin{align}
    \mathcal{F}^{2,2}_0 &= \left(\mathcal{W}^{\widehat{\mathfrak{su}}(2)_2}_{l=0}\otimes \mathcal{W}^{\widehat{\mathfrak{su}}(2)_2}_{l=0}\oplus\mathcal{W}^{\widehat{\mathfrak{su}}(2)_2}_{l=2}\otimes \mathcal{W}^{\widehat{\mathfrak{su}}(2)_2}_{l=2}\right) \otimes \mathcal{W}^{\widehat{\mathfrak{u}}(1)}_{0},\label{duality1}\\
    \mathcal{F}^{2,2}_2 &= \left(\mathcal{W}^{\widehat{\mathfrak{su}}(2)_2}_{l=0}\otimes \mathcal{W}^{\widehat{\mathfrak{su}}(2)_2}_{l=2}\oplus\mathcal{W}^{\widehat{\mathfrak{su}}(2)_2}_{l=2}\otimes \mathcal{W}^{\widehat{\mathfrak{su}}(2)_2}_{l=0}\right) \otimes \mathcal{W}^{\widehat{\mathfrak{u}}(1)}_{2}\label{duality2}
\end{align}
\paragraph{Character of the fermionic Fock space}
We use the following formula for the character of the free fermion Fock space $\mathcal{F}^{2,2}_\sigma$
\begin{equation}
    ch(\mathcal{F}^{2,2}_l) = \mbox{Res}_{p=0} p^{-\sigma-1} \prod_{n=1}^\infty \left((1+p q^{n-1/2})^4(1+p^{-1}q^{n-1/2})^4\right)
\end{equation}
where the residue picks up the contribution of fermionic number $\sigma$ in the usual free fermion partition function.
For $l=0,2$, we find at first orders in $q$,
\begin{align}
    ch(\mathcal{F}^{2,2}_0)& =1 + 16 q + 68 q^2 + 256 q^3 + 777 q^4 + 2160 q^5 + 5460 q^6 + 
 13056 q^7 + 29482 q^8+\cdots\,,\\
   ch(\mathcal{F}^{2,2}_2)&= 6 q+32 q^2+140 q^3+448 q^4+1316 q^5+3456 q^6+8520 q^7+19712 q^8+\cdots\,.
\end{align}

\paragraph{Character for $\widehat{\mathfrak{su}}(2)$ Kac-Moody:} The character formula for the $\widehat{\mathfrak{su}}(2)$ Kac-Moody algebra at level $k$ can be found in \cite{DiFrancesco1997},
\begin{equation}
    \chi^{\widehat{\mathfrak{su}}(2)}_{l,k}=\frac{1}{\eta(q)^3}\sum_{n\in \mathbb{Z}}(l+1+2(k+2)n)q^{\frac{1}{4(k+2)}(l+1+2(k+2)n)^2}
\end{equation}
where $l=0,1,\cdots,k$ labels of primary fields and 
\begin{equation}
    \eta(q)=q^{1/24}\prod_{n=1}^\infty (1-q^n)\,.
\end{equation}
In particular, removing the overall factor $q^{-c/24}=q^{-1/16}$,
\begin{align}
    ch(\mathcal{W}^{\widehat{\mathfrak{su}}(2)_2}_{l=0}) &=1+3 q+9 q^2+15 q^3+30 q^4+54 q^5+94 q^6+153 q^7+252 q^8+\cdots\,,\\
    ch(\mathcal{W}^{\widehat{\mathfrak{su}}(2)_2}_{l=0}) &=q^{1/2}\left(
    3 + 4 q + 12 q^2 + 21 q^3 + 43 q^4 + 69 q^5 + 123 q^6 + 193 q^7 + 
 318 q^8+\cdots\right)\,,
\end{align}

\paragraph{Character for $\widehat{\mathfrak{u}}(1)$ Kac-Moody:}
We use,
\begin{align}
    ch(\mathcal{W}^{\widehat{\mathfrak{u}}(1)}_{\sigma})&=q^{\sigma^2/8}\prod_{n=1}^\infty(1-q^n)^{-1}=q^{\sigma^2/8}\left(
    1+q+2 q^2+3 q^3+5 q^4+7 q^5+11 q^6+15 q^7+22 q^8+\cdots\right)\,.
\end{align}

Using these formulas, we can check the following character identities at first orders in $q$,
\begin{align}
       ch(\mathcal{F}^{2,2}_0) &=\left( ch(\mathcal{W}^{\widehat{\mathfrak{su}}(2)_2}_{l=0})^2+ch(\mathcal{W}^{\widehat{\mathfrak{su}}(2)_2}_{l=2})^2 \right)ch(\mathcal{W}^{\widehat{\mathfrak{u}}(1)}_{0})\,,\\
       ch(\mathcal{F}^{2,2}_2) &=2ch(\mathcal{W}^{\widehat{\mathfrak{su}}(2)_2}_{l=0})ch(\mathcal{W}^{\widehat{\mathfrak{su}}(2)_2}_{l=2} )ch(\mathcal{W}^{\widehat{\mathfrak{u}}(1)}_{0})\,.
\end{align}

\bibliographystyle{utphys}
\bibliography{QHMM}
\end{document}